\documentclass[floats,aps,epsf,amsfonts]{revtex4}
\usepackage[all]{xy}
\usepackage{amsmath}
\usepackage[dvips]{graphicx}
\begin{document}

\title{$m$-mode regularization scheme for the self force in Kerr spacetime}

\author{Leor Barack, Darren A Golbourn and Norichika Sago}
\affiliation
{School of Mathematics, University of Southampton, Southampton,
SO17 1BJ, United Kingdom}

\date{\today}

\begin{abstract}
We present a new, simple method for calculating the scalar, electromagnetic, and
gravitational self forces acting on particles in orbit around a Kerr black hole.
The standard ``mode-sum regularization'' approach for self-force calculations relies
on a decomposition of the full (retarded) perturbation field into multipole modes,
followed by the application of a certain mode-by-mode regularization procedure.
In recent years several groups have developed numerical codes for calculating black
hole perturbations directly in 2+1 dimensions (i.e., decomposing the azimuthal
dependence into $m$-modes, but refraining from a full multipole decomposition).
Here we formulate a practical scheme for constructing the self force directly
from the 2+1-dimensional $m$-modes. While the standard mode-sum method is
serving well in calculations of the self force in Schwarzschild geometry,
the new scheme should allow a more efficient treatment of the Kerr problem.
\end{abstract}

\maketitle

\section{Introduction}

The motion of a small particle (a point mass, a point electric charge, or a
point scalar charge) in curved spacetime, in situations where the perturbation
caused by the particle can be considered ``small'', may be described in terms
of an effective self force (SF). The SF accelerates the particle with respect
to the ``background'' spacetime; it contains both conservative and dissipative
parts, the latter being interpreted as the radiation reaction force. A formal
expression for the electromagnetic (EM) SF in curved spacetime was derived in
1960 by DeWitt and Brehme \cite{DeWitt:1960fc} (later corrected by Hobbs \cite{Hobbs:1968}).
In 1997, Mino, Sasaki and Tanaka \cite{Mino:1996nk} derived a formal expression
for the gravitational SF, in two different ways: by extending the method of DeWitt
and Breheme, and also using the technique of matched asymptotic expansions.
An alternative derivation of both EM and gravitational SFs was
presented by Quinn and Wald \cite{Quinn:1996am}. The formal expression for the
scalar SF was obtained by Quinn \cite{Quinn:2000wa} in 2000. Detweiler and
Whiting \cite{Detweiler:2002mi} later showed how the SF interpretation of the motion
is consistent with the standard picture of geodesic motion in a perturbed geometry.
A thorough self-contained review of these developments can be found in
\cite{Poisson:2003nc}.

For all fields (scalar, EM, or gravitational), the formal expression for the SF
contains a ``tail'' term, which describes the interaction of the particle
with the part of its field supported {\em inside} the particle's past light-cone.
In the gravitational case, the tail part of the self-interaction is responsible for
the entire SF effect (for particles moving along momentarily-geodesic trajectories
in vacuum spacetimes).
In practice, to facilitate actual calculations of the SF in concrete situations,
it is convenient to express the tail force as the difference between
the ``full'' force, arising from the full (retarded) perturbation field, and the
``direct'' force, describing the back reaction from the instantaneous part of
the retarded field supported only {\em on} the particle's past light-cone.

The implementation of the above theoretical framework in the concrete
problem of a particle in orbit around a black hole has been the subject of intensive
study over the last decade. (This has been largely motivated by the need to accurately
model the orbital evolution of astrophysical compact objects inspiralling into
massive black holes---of the prime targets for LISA, the planned space-based
gravitational wave detector.) To calculate the SF in this scenario, one
usually starts by solving (numerically) the appropriate linear perturbation
equations over the black hole background, with a stress-energy source
corresponding to the orbiting particle. One normally exploits the symmetry
of the black hole background, by writing the perturbation equations in a separated
form and tackling them mode by mode. The main practical challenge, then, is to
correctly split each of full, retarded-field modes into its ``tail'' and
``direct'' contributions. A scheme to achieve this was first devised by Barack
and Ori in 2000 \cite{Barack:1999wf}, and later developed to deal with arbitrary
geodesic orbits in Schwarzschild
\cite{Barack:2001bw,Barack:2001gx,Barack:2002mha,Barack:2002bt}, and, eventually,
arbitrary geodesic orbits in Kerr \cite{Barack:2002mh}. This ``mode-sum regularization''
scheme has since been implemented by various authors on a case-by case basis,
so far only for orbits in Schwarzschild
\cite{Barack:2000eh,Barack:2002ku,Burko:1999zy,Barack:2000zq,
Burko:2000xx,Detweiler:2002gi,Diaz-Rivera:2004ik,Nakano:2003he}.
Most recently, the mode-sum scheme was used to calculate
the scalar SF for generic geodesic orbits in Schwarzschild
\cite{Haas:2007kz}, and the gravitational SF for circular geodesics in
Schwarzschild \cite{Barack:2007tm}.

The standard mode-sum scheme uses as input the individual multipole modes of the
full perturbation field (which are normally obtained numerically).
The contribution of each mode to the full force (which is finite at the location of the
particle) is subjected to a certain regularization procedure, which essentially
amounts to subtracting out the direct force mode-by-mode. The sum of the regularized
modes is guaranteed to converge and yield the correct tail force.
In this description, the term ``multipole mode'' refers to a particular
spherical-harmonic component $\ell$ of the field, summed over all azimuthal numbers $m$.
In spherically-symmetric black hole spacetimes (including Schwarzschild),
these $\ell$-modes can be obtained using either frequency-domain or
time-domain methods.
In the former approach (e.g., \cite{Diaz-Rivera:2004ik}) one first solves the ODE
for each Fourier frequency $\omega$ of each multipole $\ell,m$ of the field, and
then sums the contributions from all $\omega$ and $m$ for given $\ell$.
In the second, time-domain approach (e.g., \cite{Haas:2007kz,Barack:2007tm}), one
solves a PDE in 1+1D (radius+time) for each $\ell,m$, and then
sums over $m$.

The situation in Kerr is slightly more involved, as in this case the field
equations do not separate into individual $\ell,m$ modes in the time domain,
and the 1+1D approach is not applicable. In this case, the $\ell$-modes
required as input for the mode-sum scheme are obtained by first solving
the ODE for each (spin-weighted) spheroidal-harmonic $\tilde\ell,m,\omega$
of the perturbation, then summing over $\omega,m$ for given $\tilde\ell$, and
finally re-decomposing each spheroidal-harmonic $\tilde\ell$-mode in a basis
of spherical harmonics $\ell$. Although this procedure is mathematically quite
straightforward, its implementation may be rather cumbersome. It has
not been attempted so far.

An alternative, more direct approach in the Kerr case may become possible thanks
to recent developments which facilitate the calculation of black hole perturbations in
2+1D. In this calculations one exploits only the axial symmetry of the underlying Kerr
geometry: The field equations are decomposed only into azimutal modes (each $\propto
e^{im\varphi}$, where $\varphi$ is the Boyer-Lindquist azimuthal coordinate),
and one solves directly for the 2+1D $m$-modes using time evolution. Over the past decade,
several authors have considered evolution in 2+1D, with or without a particle source.
Krivan \emph{et al.\ }\cite{Krivan:1996da,Krivan:1997hc} explored the late-time decay of scalar
fields and Weyl-scalar vacuum perturbations by evolving the master Teukolsky equation in
2+1D. Pazos-Avalos and Lousto \cite{PazosAvalos:2004rp} presented an improved,
fourth-order-convergent code in 2+1D, for the evolution of vacuum perturbations of
the Teukolsky equation. Particle orbits in Kerr were tackled in 2+1D by
Lopez-Aleman \emph{et al.\ }\cite{LopezAleman:2003ik}, Khanna \cite{Khanna:2003qv}
and Burko and Khanna \cite{Burko:2006ua}, using a Gaussian representation of the
particle. A more sophisticated finite-impulse representation was very recently
suggested and implemeted by Sundararajan \emph{et al.\ }\cite{Sundararajan:2007jg}.
Sopuerta \emph{et al.}\ proposed the use of finite-element methods for an effective
treatment of the particle in 2+1D.
This idea was implemented so far only in a 1+1D context
\cite{Sopuerta:2005rd,Sopuerta:2005gz},
and it awaits further development.
Most recently, Barack and Golbourn \cite{Barack:2007jh}
proposed a ``puncture'' scheme (further discussed below)
for dealing with the singular behavior of the field in 2+1D.
They demonstrated the applicability of this method in the
test case of a scalar charge set in a circular orbit around
a Schwarzschild black hole.

Suppose that, for a given orbital configuration in Kerr, we had at hand the
2+1D $m$-modes of the retarded perturbation field (say, in the form of numerical
solutions). We could then proceed by decomposing these modes into spherical-harmonic
$\ell,m$-modes, which, upon summation over $m$ for given $\ell$
and evaluation of the contribution from each $\ell$-mode to the full force,
would provide the necessary input for the standard mode-sum scheme.
However, it would clearly be much more straightforward to construct the
SF directly from the $m$-modes, avoiding the $\ell$-decomposition altogether.
{\it The goal of this work is to formulate such an ``$m$-mode regularization''
scheme, which allows access to the SF directly from the 2+1D $m$-modes.}

There is an important difference between $\ell$-mode and $m$-mode formulations:
The $\ell$-mode of the full, retarded field (summed over $m$ for given $\ell$)
is {\em continuous} at the particle's location, and its contribution to the full
force is finite. This is true for all types of
perturbations, scalar, EM and gravitational (in the Lorenz gauge) alike.
In contrast, the 2+1D $m$-mode perturbation {\em diverges} at the particle
(logarithmically, in general \cite{Barack:2007jh}), and its contribution to the
full force is divergent as well. This is troublesome in two ways: Firstly,
it complicates significantly the numerical treatment. A numerical evolution
scheme formulated in 2+1D, with a pointlike source, has to deal somehow with
the divergence of the solutions at the particle's location. Secondly, since the
individual $m$-mode contributions to the full force are divergent, they cannot be
used, as they are, as input for an $m$-mode regularization
scheme---at least not in the same way that the (finite, well defined) $\ell$-modes
are used as input for the standard mode-sum scheme.

The puncture method of Ref.\ \cite{Barack:2007jh} offers solutions for both above
difficulties. To illustrate the essential technique, consider the case of
scalar perturbations (the EM and gravitational cases will be described later).
Let $\Phi$ be the retarded perturbation field caused by a scalar charge
in orbit around a Kerr black hole. In the puncture scheme we formally split
$\Phi$ as
%~~~~~~~~~~~~~~~~~~~~~~~~~~~~~~~~~~~~~~~~~~~~~~~~~~~~~~~~~~~~~~~~~~~~~~
\begin{equation}\label{I10}
\Phi=\Phi_{\rm res}+\Phi_{\rm P},
%%\quad\quad\text{\P I10}
\end{equation}
%~~~~~~~~~~~~~~~~~~~~~~~~~~~~~~~~~~~~~~~~~~~~~~~~~~~~~~~~~~~~~~~~~~~~~~
where the ``puncture field'' $\Phi_{\rm P}$ is a certain function, given
analytically, which approximates the singular behavior of $\Phi$
well enough to guarantee that the $m$-modes of the ``residual field''
$\Phi_{\rm res}$ are continuous at the particle's location
\footnote{Our residual field $\Phi_{\rm res}$ is not to be confused
with the ``R-field'' introduced by Detweiler and Whiting
\cite{Detweiler:2002mi}: The latter, by definition, is a
solution of the vacuum field equation, whereas our function
$\Phi_{\rm res}$ is not necessarily so.}.
We then utilize $\Phi_{\rm res}$ as a new variable for the numerical
evolution: We re-write the field equation in terms of $\Phi_{\rm res}$
(it will now have a source term which depends on the known function $\Phi_{\rm P}$),
separate the azimuthal dependence, and use evolution in 2+1D to solve for
each of the $m$-modes of $\Phi_{\rm res}$, which are continuous fields.
%(In practice, this procedure
%is applied only in a neighborhood of the particle; the numerical evolution
%switches back to the original variable $\Phi$ away from the
%particle---see \cite{Barack:2007jh} for details.)
The $m$-mode of the actual (divergent)
field $\Phi$ is then just the sum of the $m$-mode of $\Phi_{\rm res}$,
obtained numerically, and the $m$-mode of $\Phi_{\rm P}$, given analytically.
This scheme deals with the first of the aforementioned difficulties. It does so,
essentially, by treating the singularity of the field analytically, and solving
numerically only for the residual, continuous field.

The analysis in Ref.\ \cite{Barack:2007jh} focused on the calculation of the (scalar)
field itself, and was not concerned with its derivatives. It incorporated a
``leading-order'' puncture, for which the $m$-modes of $\Phi_{\rm res}$ were
continuous but generally not differentiable (their derivatives diverged
logaritmically at the particle).
In order to deal with the second of the aforementioned complications---the
divergence of the $m$-mode contributions to the full force---we will consider in
the present work an improved version of the puncture scheme, in which the $m$-modes of
$\Phi_{\rm res}$ are not only continuous, but also have continuous derivatives at
the particle. We will prescribe a puncture function $\Phi_{\rm P}$ which
achieves this. Our new mode-sum scheme for the SF will require as input the
$m$-modes of the residual function $\Phi_{\rm res}$.
We will show that the correct tail part of the scalar SF is simply
the sum over all $m$-mode forces exerted on the particle by the
$m$-modes of $\Phi_{\rm res}$, with no further regularization required.
Similar results will apply in the EM and gravitational cases.

This paper is structured as follows. Sections II, III, and IV are each
devoted to one of the field types: Scalar, EM, and gravitational, in order.
Each of these sections contains three parts. In the first part we review
the formulation of the equation of motion with a SF term;
in the second part we prescribe our $m$-mode regularization scheme for
the SF; and in the third part we justify the scheme and explain the
derivation of the new mode-sum formula.
Section V contains a summary and some concluding remarks.

Throughout this work we use standard geometrized units (with $c=G=1$)
and metric signature $({-}{+}{+}{+})$. The Riemann tensor is defined as in
Ref.\ \cite{MTW}, and $t,r,\theta,\varphi$ are the standard Boyer-Lindquist
coordinates.

%%%%%%%%%%%%%%%%%%%%%%%%%%%%%%%%%%%%%%%%%%%%%%%%%%%%%%%%%%%%%%%%%%%%%%%%%%%%%%
\section{Scalar field} \label{Scalar}
%%%%%%%%%%%%%%%%%%%%%%%%%%%%%%%%%%%%%%%%%%%%%%%%%%%%%%%%%%%%%%%%%%%%%%%%%%%%%%

\subsection{Preliminaries}

Consider a test particle of a scalar charge $q$, moving freely
in the vacuum exterior of a Kerr black hole with mass $M\gg q$ and angular
momentum $aM$. Neglecting SF effects, the particle moves along a geodesic
$x^{\mu}=z^{\mu}(\tau)$ of the background spacetime, parameterized
by $\tau$, the proper time. We denote the full (retarded) scalar field associated
with this particle by $\Phi(x)$, and assume that it satisfies the
minimally-coupled Klein--Gordon equation,
%~~~~~~~~~~~~~~~~~~~~~~~~~~~~~~~~~~~~~~~~~~~~~~~~~~~~~~~~~~~~~~~~~~~~~~
\begin{equation}\label{S10}
\nabla^{\alpha}\nabla_{\alpha}\Phi=-4\pi\rho.
%\quad\quad\text{\P S10}
\end{equation}
%~~~~~~~~~~~~~~~~~~~~~~~~~~~~~~~~~~~~~~~~~~~~~~~~~~~~~~~~~~~~~~~~~~~~~~
Here $\nabla_{\alpha}$ denotes covariant differentiation with respect to the
background Kerr geometry, and indices are raised and lowered using the background Kerr
metric $g_{\alpha\beta}$. The scalar charge density on the right-hand side (RHS)
is given by
%~~~~~~~~~~~~~~~~~~~~~~~~~~~~~~~~~~~~~~~~~~~~~~~~~~~~~~~~~~~~~~~~~~~~~~
\begin{equation}\label{S20}
\rho(x)=q\int_{-\infty}^{\infty}\delta^4[x-z(\tau)](-g)^{-1/2}d\tau,
%\quad\quad\text{\P S20}
\end{equation}
%~~~~~~~~~~~~~~~~~~~~~~~~~~~~~~~~~~~~~~~~~~~~~~~~~~~~~~~~~~~~~~~~~~~~~~
where $g$ is the determinant of $g_{\alpha\beta}$, and $x,z$ are short-hand
for $x^{\mu},z^{\mu}$.

Including self interaction of order $q^2$ (and ignoring the gravitational SF),
the equation of motion of the particle can be written in the form \cite{Quinn:2000wa}
%~~~~~~~~~~~~~~~~~~~~~~~~~~~~~~~~~~~~~~~~~~~~~~~~~~~~~~~~~~~~~~~~~~~~~~
\begin{equation}\label{S30}
u_{\beta}\nabla^{\beta}(\mu u^{\alpha})=F_{\rm self}^{\alpha}=
\lim_{x\to z(\tau)} F_{\rm tail}^{\alpha}(x),
%\quad\quad\text{\P S30}
\end{equation}
%~~~~~~~~~~~~~~~~~~~~~~~~~~~~~~~~~~~~~~~~~~~~~~~~~~~~~~~~~~~~~~~~~~~~~~
where $\mu$ is the rest mass of the particle (assumed much smaller than $M$),
$u^{\alpha}\equiv dz^{\alpha}/d\tau$ is the particle's four-velocity,
%The two terms on the RHS describe the SF: $F_{\rm ALD}^{\alpha}$
%is a local term analogous to the Abraham-Lorentz-Dirac force of flat-space
%electrodynamics (it depends on the four-acceleration and its time derivative at
%$z(\tau)$---see \cite{Quinn:2000wa} for the explicit form].
and the ``tail force'' field is defined by
%$F_{\rm tail}^{\alpha}$ is the tail-force field , which brings the dependence
%on the particle's past, and whose calculation is the main concern of this paper.
%The tail term is obtained from a ``tail force'' field $F_{\rm tail}^{\alpha}(x)$,
%given by
%~~~~~~~~~~~~~~~~~~~~~~~~~~~~~~~~~~~~~~~~~~~~~~~~~~~~~~~~~~~~~~~~~~~~~~
\begin{equation}\label{S40}
F_{\rm tail}^{\alpha}(x)=q^2 \lim_{\epsilon\to 0^+} \int_{-\infty}
^{\tau_-(x)-\epsilon}\nabla^{\alpha}G_{\rm ret}[x,z(\tau')]d\tau'.
%\quad\quad\text{\P S40}
\end{equation}
%~~~~~~~~~~~~~~~~~~~~~~~~~~~~~~~~~~~~~~~~~~~~~~~~~~~~~~~~~~~~~~~~~~~~~~
Here $G_{\rm ret}[x,x']$ is the retarded Green's function associated with Eq.\
(\ref{S10}), $\nabla^{\alpha}$ acts on the first argument of $G_{\rm ret}$,
and $\tau_-(x)$ is the value of $\tau$ at which the past light-cone
of point $x$ intersects the particle's worldline. Notice that in Eq.\ (\ref{S30}) we have
kept $\mu$ inside the derivative. This is necessary, as $\mu$ will generally
be time-dependent (this relates to the fact that, in general, the orbiting scalar
particle will emit monopole radiation, a process which alters the particle's rest mass).
The components of Eq.\ (\ref{S30}) orthogonal and tangent to $u^{\alpha}$ give,
respectively, the acceleration of the particle and the rate
of change of $\mu$.\footnote{In this regard, we point to a mistake in Eq.\ (1)
of Ref.\ \cite{Barack:2002mha}:
This form of the scalar-particle equation of motion is,
in general, not self-consistent, since the SF, as defined in
\cite{Barack:2002mha}, is not necessarily co-alighned with the
four-acceleration. The correct form is given in Eq.\ (\ref{S30})
here.}

We comment on the regularity of the tail term, which will play an important
role in our analysis: The limit in Eq.\ (\ref{S40})
cuts the worldline integral short of the light-cone singularity of the
integrand at $z=z(\tau_-)$. As a result, the integrand is a smooth function of $x$
even at $x\to z$. Since $\tau_-(x)$ is a continuous function of $x$
[as $x\to z(\tau)$ we have $\tau_-(x)\to\tau$], the field
$F_{\rm tail}^{\alpha}(x)$ is continuous near $x=z$, and the tail term $F_{\rm
tail}^{\alpha}[z(\tau)]$ in Eq.\ (\ref{S30}) is well defined.
However, the integration limit $\tau_-(x)$ does not depend smoothly on the
coordinates $x$, which impairs the smoothness of $F_{\rm tail}^{\alpha}(x)$.
More precisely, $\nabla_{\beta}\tau_-(x)$ has a finite-jump discontinuity
at $x=z$ [see Eq.\ (16) of \cite{Quinn:2000wa}], which renders
the derivatives of $F_{\rm tail}^{\alpha}(x)$ discontinuous, yet bounded, at $x=z$.
Our basic working assumption in the formulation of the $m$-mode scheme below will be
that the tail-force field $F_{\rm tail}^{\alpha}(x)$ is continuous for all $x$,
and has at least piecewise continuous derivatives.

\subsection{$m$-mode scheme: A prescription}

In the following we let $z(\tau)$ be an arbitrary bound geodesic orbit of a scalar
particle around a Kerr black hole, and
%and we wish to calculate the SF at a particular point
%$z_0\equiv z(\tau_0)$ along this trajectory, with Boyer-Lindquist coordinates
%$z_0^{\alpha}=(t_0,r_0,\theta_0,\varphi_0)$. We will
prescribe the construction of the SF at an arbitrary point along this orbit,
using the proposed $m$-mode scheme. The basic SF construction formula is
Eq.\ (\ref{S120}) below. In this subsection we merely state this formula;
in the next subsection we will explain its derivation.

\begin{itemize}

\item {\it Step 1: Construct the puncture function.} \\

For an arbitrary spacetime point $x$ outside the black hole, let $\Sigma$ be
the spatial hypersurface $t$=const containing $x$, let $\bar\tau(t)$ be
the value of $\tau$ at which the particle's worldline is intersected by $\Sigma$, and denote
$\bar z(t)\equiv z[\bar\tau(t)]$. Define the coordinate distance
$\delta x^{\alpha}\equiv x^{\alpha}-\bar z^{\alpha}(t)$, and construct the
two quantities
%~~~~~~~~~~~~~~~~~~~~~~~~~~~~~~~~~~~~~~~~~~~~~~~~~~~~~~~~~~~~~~~~~~~~~~
\begin{equation} \label{S50}
S_0=\left. (g_{\alpha\beta}+u_{\alpha}u_{\beta})\right|_{\bar z}
\delta x^{\alpha}\delta x^{\beta},
\quad\quad
S_1=\left. \left(u_{\lambda}u_{\gamma}\Gamma^{\lambda}_{\alpha\beta}+
g_{\alpha\beta,\gamma}/2 \right)\right|_{\bar z}
\delta x^{\alpha}\delta x^{\beta}\delta x^{\gamma},
%\quad\quad\text{\P S50}
\end{equation}
%~~~~~~~~~~~~~~~~~~~~~~~~~~~~~~~~~~~~~~~~~~~~~~~~~~~~~~~~~~~~~~~~~~~~~~
where the four-velocity $u_{\alpha}$, the Kerr metric $g_{\alpha\beta}$,
its derivatives $g_{\alpha\beta,\gamma}$ and the connection coefficients
$\Gamma^{\lambda}_{\alpha\beta}$ are all evaluated at the worldline point
$\bar z$.
Finally, define the {\it puncture function} as
%~~~~~~~~~~~~~~~~~~~~~~~~~~~~~~~~~~~~~~~~~~~~~~~~~~~~~~~~~~~~~~~~~~~~~~
\begin{equation} \label{S60}
\Phi_{\rm P}=\frac{q}{\epsilon_{\rm P}}, \quad \text{where}\quad
\epsilon_{\rm P}=\sqrt{S_0+S_1}.
%\quad\quad\text{\P S60}
\end{equation}
%~~~~~~~~~~~~~~~~~~~~~~~~~~~~~~~~~~~~~~~~~~~~~~~~~~~~~~~~~~~~~~~~~~~~~~

{\em For $\delta x^{\gamma}$ very small} we have
$S=S_0+S_1+O(\delta x^4)$, where $S$ is the squared geodesic distance from
$x$ to the particle's worldline (i.e., the squared length of the small spatial geodesic section
connecting $x$ to the worldline and normal to it). The above two leading terms
of $S$, $S_0(\propto\delta x^2)$ and $S_1(\propto\delta x^3)$, are derived, e.g., in
Appendix A of Ref.\ \cite{Barack:2002mha}. We emphasize, however, that here the definitions
in Eqs.\ (\ref{S50}) and (\ref{S60}) apply for arbitrary $\delta x$, not necessarily small.

\item {\it Step 2: Write down the field equation for $\Phi_{\rm res}$ and separate
into $m$-modes.} \\

Define the {\it residual field}
%~~~~~~~~~~~~~~~~~~~~~~~~~~~~~~~~~~~~~~~~~~~~~~~~~~~~~~~~~~~~~~~~~~~~~~
\begin{equation} \label{S70}
\Phi_{\rm res}=\Phi-\Phi_{\rm P}
%\quad\quad\text{\P S70}
\end{equation}
%~~~~~~~~~~~~~~~~~~~~~~~~~~~~~~~~~~~~~~~~~~~~~~~~~~~~~~~~~~~~~~~~~~~~~~
as in Eq.\ (\ref{I10}), and re-write the scalar field equation (\ref{S10})
in the form
%~~~~~~~~~~~~~~~~~~~~~~~~~~~~~~~~~~~~~~~~~~~~~~~~~~~~~~~~~~~~~~~~~~~~~~
\begin{equation}\label{S80}
\nabla^{\alpha}\nabla_{\alpha}\Phi_{\rm res}=
-4\pi\rho-\nabla^{\alpha}\nabla_{\alpha}\Phi_{\rm P}\equiv Z_{\rm res}.
%\quad\quad\text{\P S80}
\end{equation}
%~~~~~~~~~~~~~~~~~~~~~~~~~~~~~~~~~~~~~~~~~~~~~~~~~~~~~~~~~~~~~~~~~~~~~~
The source $Z_{\rm res}$ is extended (not confined to the particle's worldline), but
contains no Dirac-delta on the worldline. As we show later,
the field $\Phi_{\rm res}$ is continuous at the particle, and its
derivatives there are bounded (albeit generally discontinuous).

Now formally decompose $\Phi_{\rm res}$ and $Z_{\rm res}$ into azimuthal $m$-modes,
in the form
%~~~~~~~~~~~~~~~~~~~~~~~~~~~~~~~~~~~~~~~~~~~~~~~~~~~~~~~~~~~~~~~~~~~~~~
\begin{equation}\label{S90}
\Phi_{\rm res}=\sum_{m=-\infty}^{\infty} \phi^m_{\rm res}(t,r,\theta) e^{i m \varphi},
\quad\quad
Z_{\rm res}=\sum_{m=-\infty}^{\infty} Z^m_{\rm res}(t,r,\theta) e^{i m \varphi},
%\quad\quad\text{\P S80}
\end{equation}
%~~~~~~~~~~~~~~~~~~~~~~~~~~~~~~~~~~~~~~~~~~~~~~~~~~~~~~~~~~~~~~~~~~~~~~
and use these expansions to separate the $\varphi$ dependence in Eq.\
(\ref{S80}). Each of the (complex-valued) $m$-modes $\phi^m_{\rm res}(t,r,\theta)$
will satisfy a hyperbolic field equation in 2+1D, of the form
%~~~~~~~~~~~~~~~~~~~~~~~~~~~~~~~~~~~~~~~~~~~~~~~~~~~~~~~~~~~~~~~~~~~~~~
\begin{equation}\label{S100}
\Box^{(3)}_{\rm S}\phi_{\rm res}^m= Z_{\rm res}^m,
%\quad\quad\text{\P S100}
\end{equation}
%~~~~~~~~~~~~~~~~~~~~~~~~~~~~~~~~~~~~~~~~~~~~~~~~~~~~~~~~~~~~~~~~~~~~~~
where $\Box^{(3)}_{\rm S}$ is a certain ($m$-dependent) second-order differential
operator. The source modes are given explicitly by\footnote{
Throughout this work we take the principal values of the Boyer-Lindquist
azimuthal coordinate to lie in the range $-\pi<\varphi\leq \pi$.}
%~~~~~~~~~~~~~~~~~~~~~~~~~~~~~~~~~~~~~~~~~~~~~~~~~~~~~~~~~~~~~~~~~~~~~~
\begin{equation}\label{S110}
Z^m_{\rm res}=\frac{1}{2\pi}\int_{-\pi}^{\pi}Z_{\rm res}
e^{-im\varphi'}d\varphi',
%\quad\quad\text{\P S110}
\end{equation}
%~~~~~~~~~~~~~~~~~~~~~~~~~~~~~~~~~~~~~~~~~~~~~~~~~~~~~~~~~~~~~~~~~~~~~~
which can be evaluated either analytically (as in \cite{Barack:2007jh}) or numerically.
As we show later, the $m$-modes $\phi^m_{\rm res}$ are continuous and differentiable
(have continuous first derivatives) at the particle.

\item {\it Step 3: Formulate an initial/boundary-condition problem for
$\phi_{\rm res}^m$}. \\

We are looking for particular solutions $\phi_{\rm res}^m$ which, through
$\phi_{\rm res}^m+\phi_{\rm P}^m=\phi^m$, give the physical,
retarded-field modes $\phi^m$. [Here $\phi_{\rm P}^m$ and
$\phi^m$ are the $m$-modes of $\Phi_{\rm P}$ and $\Phi$,
defined in analogy to $\phi_{\rm res}^m$ in Eq.\ (\ref{S90})].
For this to happen, the boundary conditions for $\phi_{\rm res}^m$ should be
such that $\phi_{\rm res}^m+\phi_{\rm P}^m$ represents purely outgoing radiation
at the far ``wave zone'' ($t,r\gg M$), and purely ingoing radiation
near the event horizon. Since the puncture modes $\phi_{\rm P}^m$ are given
analytically (at least in terms of closed-form definite integrals), this
readily translates to physical boundary conditions for $\phi_{\rm res}^m$.
In practice, it may be more convenient to re-define the puncture $\phi_{\rm P}$
by (smoothly) suppressing its support away from the particle. The new residual
function $\phi_{\rm res}$ then coincides with $\phi$ away from the
particle, and the usual ingoing/outgoind boundary conditions will apply to it.
A similar idea was implemented in the analysis of Ref.\ \cite{Barack:2007jh}.

The choice of correct initial conditions for the numerical evolution is less
of a concern. Linear perturbations on black hole backgrounds can be evolved
stably over an indefinite amount of time. If the initial data for the evolution are
specified much earlier than the time along the orbit where we wish to determine the
SF, then the exact form of initial data (chosen reasonably enough) will have little effect
on the outcome of our calculation. This, too, is demonstrated in the analysis
of Ref.\ \cite{Barack:2007jh}.

\item {\it Step 4: Solve for $\phi_{\rm res}^m$ using numerical evolution in 2+1D.} \\

We envisage solving Eq.\ (\ref{S100}) using a finite-difference time-evolution
code formulated in 2+1D, as in Ref.\ \cite{Barack:2007jh}. Here we shall not
be further concerned with the details of the numerical implementation, and
proceed by assuming that the modes $\phi_{\rm res}^m$ have been calculated.

\item {\it Step 5: Apply the $m$-mode-sum formula.} \\

The SF is simply given by
%~~~~~~~~~~~~~~~~~~~~~~~~~~~~~~~~~~~~~~~~~~~~~~~~~~~~~~~~~~~~~~~~~~~~~~
\begin{equation}\label{S120}
F_{\rm self}^{\alpha}[z(\tau)]=q\sum_{m=0}^{\infty}
\left.\nabla^{\alpha}\tilde\phi_{\rm res}^m \right|_{x=z(\tau)},
%\quad\quad\text{\P S120}
\end{equation}
%~~~~~~~~~~~~~~~~~~~~~~~~~~~~~~~~~~~~~~~~~~~~~~~~~~~~~~~~~~~~~~~~~~~~~~
where $\tilde\phi_{\rm res}^m(t,r,\theta,\varphi)$ are real fields constructed
from $\phi_{\rm res}^m(t,r,\theta)$ through
%~~~~~~~~~~~~~~~~~~~~~~~~~~~~~~~~~~~~~~~~~~~~~~~~~~~~~~~~~~~~~~~~~~~~~~
\begin{equation}\label{S130}
\tilde\phi_{\rm res}^{m}=
2 {\rm Re}\left(\phi_{\rm res}^m e^{im\varphi}\right)\ \text{for $m>0$},
\quad\text{and}\quad \tilde\phi_{\rm res}^{m=0}=\phi_{\rm res}^{m=0}.
%\quad\quad\text{\P S130}
\end{equation}
%~~~~~~~~~~~~~~~~~~~~~~~~~~~~~~~~~~~~~~~~~~~~~~~~~~~~~~~~~~~~~~~~~~~~~~
As we argue below, the sum in Eq.\ (\ref{S120}) is expected to converge at
least as $\sim 1/m$.

\end{itemize}

\subsection{Analysis} \label{Analysis}

In what follows we justify and explain the above mode-sum prescription.
We start by giving a detailed derivation of the mode-sum formula (\ref{S120});
we then explain its predicted convergence rate; and, finally, we analyze
the behavior of the numerical integration variables $\phi_{\rm res}^m$ near
the particle.

\subsubsection{Derivation of the mode-sum formula (\ref{S120})}

Consider the {\it tail part} of the scalar field, defined as
%~~~~~~~~~~~~~~~~~~~~~~~~~~~~~~~~~~~~~~~~~~~~~~~~~~~~~~~~~~~~~~~~~~~~~~
\begin{equation}\label{S140}
\Phi_{\rm tail}(x)=q \lim_{\epsilon\to 0^+} \int_{-\infty}
^{\tau_-(x)-\epsilon}G_{\rm ret}[x,z(\tau')]d\tau'.
%\quad\quad\text{\P S140}
\end{equation}
%~~~~~~~~~~~~~~~~~~~~~~~~~~~~~~~~~~~~~~~~~~~~~~~~~~~~~~~~~~~~~~~~~~~~~~
Recalling Eq.\ (\ref{S40}), we have
%~~~~~~~~~~~~~~~~~~~~~~~~~~~~~~~~~~~~~~~~~~~~~~~~~~~~~~~~~~~~~~~~~~~~~~
\begin{equation}\label{S150}
F_{\rm tail}^{\alpha}(x)=q\nabla^{\alpha}\Phi_{\rm tail}(x)-
q^2 (\nabla^{\alpha}\tau_-)G_{\rm ret}[x,z(\tau_-^-)],
%\quad\quad\text{\P S150}
\end{equation}
%~~~~~~~~~~~~~~~~~~~~~~~~~~~~~~~~~~~~~~~~~~~~~~~~~~~~~~~~~~~~~~~~~~~~~~
where $G_{\rm ret}[x,z(\tau_-^-)]\equiv \lim_{\epsilon\to 0^+}
G_{\rm ret}[x,z(\tau_--\epsilon)]$ contains only the smooth, tail part
of the Green's function. Examine the second term on the RHS
at the limit $x\to z$: By Eq.\ (33) of \cite{Quinn:2000wa}, we have
$\lim_{x\to z}G_{\rm ret}[x,z(\tau_-^-)]=R/12$, where $R$ is the background
scalar curvature.
Since $R=0$ for Kerr, and since, as we mentioned already, the factor
$\nabla^{\alpha}\tau_-$ is bounded at $x\to z$, we find that the
second term on the RHS of Eq.\ (\ref{S150}) vanishes at this limit. Hence,
%~~~~~~~~~~~~~~~~~~~~~~~~~~~~~~~~~~~~~~~~~~~~~~~~~~~~~~~~~~~~~~~~~~~~~~
\begin{equation}\label{S160}
F_{\rm tail}^{\alpha}=q\nabla^{\alpha}\Phi_{\rm tail}
\quad\text{for $x\to z$}.
%\quad\quad\text{\P S160}
\end{equation}
%~~~~~~~~~~~~~~~~~~~~~~~~~~~~~~~~~~~~~~~~~~~~~~~~~~~~~~~~~~~~~~~~~~~~~~

Next consider the {\it direct part} of the scalar field, defined as
%~~~~~~~~~~~~~~~~~~~~~~~~~~~~~~~~~~~~~~~~~~~~~~~~~~~~~~~~~~~~~~~~~~~~~~
\begin{equation}\label{S170}
\Phi_{\rm dir}\equiv \Phi-\Phi_{\rm tail}
=q \lim_{\epsilon\to 0^+} \int_{\tau_-(x)-\epsilon}
^{\tau_-(x)+\epsilon}G_{\rm ret}[x,z(\tau')]d\tau',
%\quad\quad\text{\P S170}
\end{equation}
%~~~~~~~~~~~~~~~~~~~~~~~~~~~~~~~~~~~~~~~~~~~~~~~~~~~~~~~~~~~~~~~~~~~~~~
and define the {\it direct force} $F_{\rm dir}^{\alpha}(x)$ by replacing
$\int_{-\infty}^{\tau_-(x)-\epsilon} \to
\int_{\tau_-(x)-\epsilon}^{\tau_-(x)+\epsilon}$ in Eq. (\ref{S40}).
We have
%~~~~~~~~~~~~~~~~~~~~~~~~~~~~~~~~~~~~~~~~~~~~~~~~~~~~~~~~~~~~~~~~~~~~~~
\begin{equation}\label{S180}
F_{\rm dir}^{\alpha}(x)=q\nabla^{\alpha}\Phi_{\rm dir}(x)+
q^2 (\nabla^{\alpha}\tau_-)G_{\rm ret}[x,z(\tau_-^-)],
%\quad\quad\text{\P S180}
\end{equation}
%~~~~~~~~~~~~~~~~~~~~~~~~~~~~~~~~~~~~~~~~~~~~~~~~~~~~~~~~~~~~~~~~~~~~~~
since $\lim_{\epsilon\to 0^+} G_{\rm ret}[x,z(\tau_-+\epsilon)]=0$ by virtue
of the retardation of the Green's function. Hence, for the direct field too,
we find
%~~~~~~~~~~~~~~~~~~~~~~~~~~~~~~~~~~~~~~~~~~~~~~~~~~~~~~~~~~~~~~~~~~~~~~
\begin{equation}\label{S190}
F_{\rm dir}^{\alpha}=q\nabla^{\alpha}\Phi_{\rm dir}
\quad\text{for $x\to z$}.
%\quad\quad\text{\P S190}
\end{equation}
%~~~~~~~~~~~~~~~~~~~~~~~~~~~~~~~~~~~~~~~~~~~~~~~~~~~~~~~~~~~~~~~~~~~~~~
Finally, define the {\it full force} $F^{\alpha}(x)$ by replacing
$\int_{-\infty}^{\tau_-(x)-\epsilon} \to \int_{-\infty}^{\tau_-(x)+\epsilon}$
in Eq. (\ref{S40}) [or, equivalently, through $F^{\alpha}(x)\equiv
F_{\rm tail}^{\alpha}(x)+F_{\rm dir}^{\alpha}(x)$].
Note $F^{\alpha}(x)=q\nabla^{\alpha}\Phi(x)$ holds
precisely, for all $x$. Hence, defining
%~~~~~~~~~~~~~~~~~~~~~~~~~~~~~~~~~~~~~~~~~~~~~~~~~~~~~~~~~~~~~~~~~~~~~~
\begin{equation}\label{S200}
F_{\rm res}^{\alpha}\equiv q\nabla^{\alpha}\Phi_{\rm res},
\quad\quad
F_{\rm P}^{\alpha}\equiv q\nabla^{\alpha}\Phi_{\rm P},
%\quad\quad\text{\P S200}
\end{equation}
%~~~~~~~~~~~~~~~~~~~~~~~~~~~~~~~~~~~~~~~~~~~~~~~~~~~~~~~~~~~~~~~~~~~~~~
we have, for all $x$,
%~~~~~~~~~~~~~~~~~~~~~~~~~~~~~~~~~~~~~~~~~~~~~~~~~~~~~~~~~~~~~~~~~~~~~~
\begin{equation}\label{S210}
F_{\rm tail}^{\alpha}(x)+F_{\rm dir}^{\alpha}(x)=
F^{\alpha}(x)=F_{\rm res}^{\alpha}(x)+F_{\rm P}^{\alpha}(x).
%\quad\quad\text{\P S210}
\end{equation}
%~~~~~~~~~~~~~~~~~~~~~~~~~~~~~~~~~~~~~~~~~~~~~~~~~~~~~~~~~~~~~~~~~~~~~~
The goal of the somewhat elaborate construction in the last tew paragraphs is,
partly, to establish the relations between the various `forces' $F(x)$ and
their corresponding fields $\Phi(x)$---these will be needed in what follows.

Based on Eq.\ (\ref{S210}) we now write the scalar SF as
%~~~~~~~~~~~~~~~~~~~~~~~~~~~~~~~~~~~~~~~~~~~~~~~~~~~~~~~~~~~~~~~~~~~~~~
\begin{eqnarray}\label{S220}
F_{\rm self}^{\alpha}(z)&=&
\lim_{x\to z} F_{\rm tail}^{\alpha}(x)
\nonumber\\
&=&
%\lim_{x\to z}\left[F^{\alpha}(x)-F_{\rm dir}^{\alpha}(x)\right]
%\nonumber\\
%&=&
\lim_{x\to z}\left[F_{\rm res}^{\alpha}(x)-\left(F_{\rm dir}^{\alpha}(x)
-F_{\rm P}^{\alpha}(x)\right)\right]
\nonumber\\
&=&
\lim_{x\to z} \sum_{m=-\infty}^{\infty}
\left[f_{\rm res}^{\alpha m}(x)-\left(f_{\rm dir}^{\alpha m}(x)
-f_{\rm P}^{\alpha m}(x)\right)\right].
%\quad\quad\text{\P S220}
\end{eqnarray}
%~~~~~~~~~~~~~~~~~~~~~~~~~~~~~~~~~~~~~~~~~~~~~~~~~~~~~~~~~~~~~~~~~~~~~~
In the last step we have formally decomposed the tail force $F_{\rm tail}^{\alpha}(x)$
into $m$-modes, introducing the notation
%~~~~~~~~~~~~~~~~~~~~~~~~~~~~~~~~~~~~~~~~~~~~~~~~~~~~~~~~~~~~~~~~~~~~~~
\begin{equation}\label{S230}
f^{\alpha m}_{\rm X}(x)\equiv\frac{1}{2\pi}\int_{-\pi}^{\pi}F^{\alpha}_{\rm
X}(y,\varphi') e^{im(\varphi-\varphi')}d\varphi',
%\quad\quad\text{\P S230}
\end{equation}
%~~~~~~~~~~~~~~~~~~~~~~~~~~~~~~~~~~~~~~~~~~~~~~~~~~~~~~~~~~~~~~~~~~~~~~
where `X' can stand for `res', `dir', or `P', and $y$ represents $t,r,\theta$.
The $m$ decomposition of $F_{\rm tail}^{\alpha}(y,\varphi)$ is technically a standard
Fourier series over the interval $-\pi<\varphi\leq \pi$ (for any fixed $y$).
Since $F_{\rm tail}^{\alpha}(x)$ is continuous and piecewise differentiable over
this interval, standard theorem of Fourier analysis (e.g., Theorem 4.2 of
\cite{James}) guarantees that the sum over $m$ in Eq.\ (\ref{S220}) convergence
to $F_{\rm tail}^{\alpha}(x)$, even at $x=z$. Moreover, the continuity of
$F_{\rm tail}^{\alpha}(x)$ assures that the Fourier sum converges {\it uniformly}
(e.g., Theorem 4.4 of \cite{James}), which allows us to switch the order of
limit and summation in Eq.\ (\ref{S220}):
%~~~~~~~~~~~~~~~~~~~~~~~~~~~~~~~~~~~~~~~~~~~~~~~~~~~~~~~~~~~~~~~~~~~~~~
\begin{equation}\label{S240}
F_{\rm self}^{\alpha}(z)=
\sum_{m=0}^{\infty} \lim_{x\to z}
\left[\tilde f_{\rm res}^{\alpha m}(x)-\left(\tilde f_{\rm dir}^{\alpha m}(x)
-\tilde f_{\rm P}^{\alpha m}(x)\right)\right].
%\quad\quad\text{\P S240}
\end{equation}
%~~~~~~~~~~~~~~~~~~~~~~~~~~~~~~~~~~~~~~~~~~~~~~~~~~~~~~~~~~~~~~~~~~~~~~
Here we have also folded the terms $m<0$ over onto $m>0$, denoting
$\tilde f_{\rm X}^{\alpha m}\equiv f_{\rm X}^{\alpha m}+f_{\rm X}^{\alpha,-m}$
for $m>0$, with $\tilde f_{\rm X}^{\alpha,m=0}\equiv f_{\rm X}^{\alpha,m=0}$.
Note that $\tilde f_{\rm X}^{\alpha m}$ are real-valued, unlike
$f_{\rm X}^{\alpha m}$ which are complex (for $m\ne 0)$.
We stress that, {\it ab initio}, the convergence of the individual sums over
$f_{\rm res}^{\alpha m}$, $f_{\rm dir}^{\alpha m}$, or $f_{\rm P}^{\alpha m}$
in Eq.\ (\ref{S220}) is not at all guaranteed. Also, one should not attempt
to apply the limit in Eq.\ (\ref{S240}) to any of the three terms
$\tilde f_{\rm X}^{\alpha m}$ individually.

We have now reached the crucial step of our derivation. In what follows we
establish that, for any $m\geq 0$,
%~~~~~~~~~~~~~~~~~~~~~~~~~~~~~~~~~~~~~~~~~~~~~~~~~~~~~~~~~~~~~~~~~~~~~~
\begin{equation}\label{S250}
{\tilde{\cal L}^{\alpha m}}\equiv \lim_{x\to z}\left(\tilde f_{\rm dir}^{\alpha m}(x)
-\tilde f_{\rm P}^{\alpha m}(x)\right)=0.
%\quad\quad\text{\P S250}
\end{equation}
%~~~~~~~~~~~~~~~~~~~~~~~~~~~~~~~~~~~~~~~~~~~~~~~~~~~~~~~~~~~~~~~~~~~~~~
To show this, we start by inspecting the difference $\Phi_{\rm dir}-\Phi_{\rm P}$
as $x\to z$. The explicit form of $\Phi_{\rm dir}$ was first worked out by Mino
{\it et al.} \cite{Mino:1998gp,Mino:2001mq}. It can be written as \cite{Barack:2002mha}
%~~~~~~~~~~~~~~~~~~~~~~~~~~~~~~~~~~~~~~~~~~~~~~~~~~~~~~~~~~~~~~~~~~~~~~
\begin{equation}\label{S260}
\Phi_{\rm dir}(x)=\frac{q}{\epsilon(x)}+\frac{qw(x)}{\epsilon(x)}+{\rm const}
\quad \text{(for $x$ near $z$)},
%\quad\quad\text{\P S260}
\end{equation}
%~~~~~~~~~~~~~~~~~~~~~~~~~~~~~~~~~~~~~~~~~~~~~~~~~~~~~~~~~~~~~~~~~~~~~~
%[see Eq.\ (17) of \cite{Barack:2002mha}],
where $\epsilon\equiv S^{1/2}$ is the geodesic distance between
$x$ and the particle's worldline, and $w$ is a certain regular function of
$x$ (and $z$), which has the form $w=O(\delta x^2)$ near the particle.
Recall $\delta x$ is short-hand for the coordinate difference
$\delta x^{\alpha}= x^{\alpha}-\bar z^{\alpha}(t)$, where $\bar z^{\alpha}$
is a point on the worldline with the same $t$ coordinate as $x$ [hence,
$\delta x^{\alpha}=(0,\delta r,\delta\theta,\delta\varphi)$].
To write the asymptotic form of $\Phi_{\rm P}$ near the particle,
we first formally expand $S$
near $\delta x=0$ as $S=S_0+S_1+S_2+\cdots$, where $S_n$ represents the
term of homogeneous order $O(\delta x^{n+2})$. [The explicit form of $S_0$
and $S_1$ was given in Eq.\ (\ref{S50}) above.] Starting from the definition
of $\Phi_{\rm P}$ in Eq.\ (\ref{S60}) we then have, for small $\delta x$,
%~~~~~~~~~~~~~~~~~~~~~~~~~~~~~~~~~~~~~~~~~~~~~~~~~~~~~~~~~~~~~~~~~~~~~~
\begin{equation}\label{S270}
\Phi_{\rm P}=\frac{q}{(S_0+S_1)^{1/2}}=
\frac{q}{\epsilon(1-S_2/S)^{1/2}}+O(\delta x^2)=
\frac{q}{\epsilon}+\frac{q S_2}{2\epsilon^3}+O(\delta x^2).
%\quad\quad\text{\P S270}
\end{equation}
%~~~~~~~~~~~~~~~~~~~~~~~~~~~~~~~~~~~~~~~~~~~~~~~~~~~~~~~~~~~~~~~~~~~~~~
From Eqs.\ (\ref{S260}) and (\ref{S270}) we obtain
%~~~~~~~~~~~~~~~~~~~~~~~~~~~~~~~~~~~~~~~~~~~~~~~~~~~~~~~~~~~~~~~~~~~~~~
\begin{equation}\label{S280}
\Phi_{\rm dir}-\Phi_{\rm P}=\frac{q\left(\epsilon^2 w-S_2/2\right)}{\epsilon^3}
+{\rm const}+O(\delta x^2).
%\quad\quad\text{\P S280}
\end{equation}
%~~~~~~~~~~~~~~~~~~~~~~~~~~~~~~~~~~~~~~~~~~~~~~~~~~~~~~~~~~~~~~~~~~~~~~
Note that both $\epsilon^2 w$ and $S_2/2$ are $O(\delta x^4)$.
Omitting the constant and the $O(\delta x^2)$ term (neither can
contribute to the eventual force at the limit $\delta x\to 0$), we can
therefore write the last expression in the form
%~~~~~~~~~~~~~~~~~~~~~~~~~~~~~~~~~~~~~~~~~~~~~~~~~~~~~~~~~~~~~~~~~~~~~~
\begin{equation}\label{S290}
\Phi_{\rm dir}-\Phi_{\rm P}=\frac{P_{[4]}(\delta x)}{\epsilon_0^3},
%\quad\quad\text{\P S290}
\end{equation}
%~~~~~~~~~~~~~~~~~~~~~~~~~~~~~~~~~~~~~~~~~~~~~~~~~~~~~~~~~~~~~~~~~~~~~~
where $\epsilon_0\equiv S_0^{1/2}$, and $P_{[4]}$ is some
multilinear function of the coordinate differences $\delta x$, of homogeneous
order $O(\delta x^4)$.
Recalling now Eqs.\ (\ref{S190}) and (\ref{S200}), we finally obtain,
at leading order in $\delta x$,
%(up to terms that vanish at $\delta x\to 0$)
%~~~~~~~~~~~~~~~~~~~~~~~~~~~~~~~~~~~~~~~~~~~~~~~~~~~~~~~~~~~~~~~~~~~~~~
\begin{equation}\label{S300}
F^{\alpha}_{\rm dir}-F^{\alpha}_{\rm P}
= q\nabla^{\alpha}\left(\Phi_{\rm dir}-\Phi_{\rm P}\right)
=\frac{q\epsilon_0^2\nabla^{\alpha}P_{[4]}-3qP_{[4]}\epsilon_0\nabla^{\alpha}\epsilon_0}
{\epsilon_0^5}
\equiv\frac{P_{[5]}^{\alpha}(\delta x)}{\epsilon_0^5},
%\quad\quad\text{\P S300}
\end{equation}
%~~~~~~~~~~~~~~~~~~~~~~~~~~~~~~~~~~~~~~~~~~~~~~~~~~~~~~~~~~~~~~~~~~~~~~
where $P_{[5]}^{\alpha}$ is yet another multilinear function of $\delta x$,
of homogeneous order $O(\delta x^5)$.
Note that $\Phi_{\rm dir}-\Phi_{\rm P}$ is continuous at the particle, and that
$F^{\alpha}_{\rm dir}-F^{\alpha}_{\rm P}$ is bounded there, yet discontinuous
(direction-dependent).

%We remind that our goal is to show the validity of Eq.\ (\ref{S250}),
%${\tilde{\cal L}^{\alpha m}}=0$.
%For this, we need to consider the $m$ decomposition of Eq.\ (\ref{S300}),
%and show that each of the $m$-modes in this decomposition (summed over
%$\pm m$ for given $m$) vanishes at the limit $\delta x\to 0$.
%A rigorous proof of this statement, through explicit evaluation of
%the relevant Fourier integral, is provided in the Appendix.
%It is instructive to give here a more heuristic argument for
%${\tilde{\cal L}^{\alpha m}}=0$, based on standard Fourier theory:
%...
%For a more rigorous proof, see the Appendix.

Our goal, recall, is to confirm the validity of Eq.\ (\ref{S250}),
i.e., show that ${\tilde{\cal L}^{\alpha m}}=0$. For this, we need to consider
the $m$ decomposition of Eq.\ (\ref{S300}), at the limit $x\to z$.
From Eqs.\ (\ref{S230}) and (\ref{S250}) we obtain
%~~~~~~~~~~~~~~~~~~~~~~~~~~~~~~~~~~~~~~~~~~~~~~~~~~~~~~~~~~~~~~~~~~~~~~
\begin{equation}\label{S310}
{\tilde{\cal L}^{\alpha m}}
=\lim_{x\to z} \frac{\alpha_m}{2\pi}\int_{-\pi}^{\pi}
\epsilon_0^{-5}(\delta y,\varphi') P_{[5]}^{\alpha}(\delta y,\varphi')
\cos[m(\varphi-\varphi')]d\varphi',
%\quad\quad\text{\P S310}
\end{equation}
%~~~~~~~~~~~~~~~~~~~~~~~~~~~~~~~~~~~~~~~~~~~~~~~~~~~~~~~~~~~~~~~~~~~~~~
where $\alpha_{m>0}=2$, $\alpha_0=1$, and $\delta y$ stands for
$\delta r,\delta\theta$ (recall $\delta t=0$). For later convenience, but
without loss of generality, we have taken here the evaluation point $z$ to
be at $\varphi=0$, so that $\delta\varphi'=\varphi'$. Note that, according to
the definition of $\bar z$, taking the limit $x\to z$ results in taking
$x\to \bar z$, and hence also $\delta x\to 0$ (i.e., $\delta y\to 0$ as well
as $\varphi\to 0$).
Inspecting the integrand in the last equation, we remind that $\epsilon_0^2$
is a positive quadratic function of $\delta y$ and $\varphi'$ which vanishes only at
the particle, and $P_{[5]}^{\alpha}$ is a sum of a terms of the form
$a^{\alpha}_{kn}\varphi'^k \delta r^n \delta\theta^{5-k-n}$, where $k,n$ are non-negative
integers satisfying $0\leq k+n\leq 5$, and $a_{kn}^{\alpha}$ are constant coefficients
(depending on $\bar z$ but not on $r,\theta$). Below we analyze separately the
terms $k<5$ and $k=5$, showing that both contributions to $\tilde{\cal L}^{\alpha m}$
vanish.

{\it Terms $k<5$:}  We define
$R(\delta y)\equiv M^{-1}\epsilon_0(\delta\varphi=0)$,
%(so that $R$ is the projection of $\epsilon_0$ onto the the $r$--$\theta$ plane)
and, for fixed values of $\delta y$ (with small $|\delta y|$), split the
integral in Eq.\ (\ref{S310}) into two domains: (i) $|\varphi'|<R$ and (ii)
$R\leq|\varphi'|<\pi$. In domain (i) we can bound $|P_{[5]}^{\alpha}|\leq c_1^{\alpha} R^5$
and $\epsilon_0\geq c_2R$, where $c_1^{\alpha}$ and $c_2$ are some positive constants
(depending on $\bar z$ but not on $r,\theta$). The absolute value of the integrand in
Eq.\ (\ref{S310}) is thus bounded from above by $c_1^{\alpha}/c_2^5$, and so the absolute
value of the integral piece $\int_{-R}^{R}$ is $\leq 2(c_1/c_2^5)R$.
Since this vanishes at the limit $\delta x\to 0$ (corresponding to $R\to 0$),
we conclude that the contribution to $\tilde{\cal L}^{\alpha m}$ from domain (i) vanishes.
Next consider domain (ii): Here we use
$\epsilon_0\geq c_3|\varphi'|$, $|\delta r|\leq c_4 R$ and $|\delta\theta|\leq c_5 R$
(where $c_3$, $c_4$ and $c_5$ are some other positive constants), to bound each
of the above $k,n$ terms of the integrand as
$|a^{\alpha}_{kn}\epsilon_0^{-5} \varphi'^k \delta r^n \delta\theta^{5-k-n}
\cos[m(\varphi-\varphi')]|\leq c_3^{-5}c_4^n c_5^{5-k-n}|a^{\alpha}_{kn}\varphi'^{k-5} R^{5-k}|$.
The absolute value of the integral over domain
(ii) is thus bounded from above by $c^{\alpha}_6\left|R^{5-k}(\pi^{k-4}-R^{k-4})\right|$
for $k<4$ and by $c^{\alpha}_7\left|R\log(\pi/R)\right|$ for $k=4$, where $c^{\alpha}_6$
and $c^{\alpha}_7$ are some positive constants. In both cases, the upper bound tends to
zero as $R\to 0$, and we conclude that the domain (ii), too, gives a null contribution to
$\tilde{\cal L}^{\alpha m}$.

{\it Term $k=5$:}
The above upper-bound argument fails when $k=5$. We now show that, nevertheless,
the contribution to $\tilde{\cal L}^{\alpha m}$ from this term vanishes as well,
this time due to the symmetry of the integrand. For $k=5$ ($\Rightarrow n=0$),
the integral in Eq.\ (\ref{S310}) takes the form
%~~~~~~~~~~~~~~~~~~~~~~~~~~~~~~~~~~~~~~~~~~~~~~~~~~~~~~~~~~~~~~~~~~~~~~
\begin{equation}\label{S320}
a^{\alpha}_{5,0}\int_{-\pi}^{\pi}
\frac{(\varphi')^5}{\epsilon_0^{5}(\delta y,\varphi')}
\cos[m(\varphi-\varphi')]d\varphi'.
%\quad\quad\text{\P S320}
\end{equation}
%~~~~~~~~~~~~~~~~~~~~~~~~~~~~~~~~~~~~~~~~~~~~~~~~~~~~~~~~~~~~~~~~~~~~~~
Once again, we split the integral into two domains, this time defined as
(i) $|\varphi'|<R^{1/2}$ and (ii) $R^{1/2}\leq|\varphi'|<\pi$.
In domain (i), the absolute value of the integrand in Eq.\ (\ref{S320}) is
bounded from above by $c_3^{-5}$, by virtue of $\epsilon_0\geq c_3|\varphi'|$.
Hence the absolute value of the integral piece $\int_{-\sqrt{R}}^{\sqrt{R}}$
is bounded from above by $2c^{-5}R^{1/2}$, which vanishes as $R\to 0$.
Thus, the contribution to $\tilde{\cal L}^{\alpha m}$
from domain (i) vanishes. Next consider domain (ii): For very small $|\delta y|$
and $|\varphi|$ we have here $|\varphi'|\geq R^{1/2}\geq |\delta r/c_4|^{1/2}
\gg |\delta r|/c_4$, and similarly $|\varphi'|\gg |\delta\theta|/c_5$ and
$|\varphi'|\gg |\varphi|$. This allows us to expand the integrand in Eq.\
(\ref{S320}) about $\delta y,\varphi=0$, in the form
%~~~~~~~~~~~~~~~~~~~~~~~~~~~~~~~~~~~~~~~~~~~~~~~~~~~~~~~~~~~~~~~~~~~~~~
\begin{equation}\label{S330}
\frac{(\varphi')^5}{\epsilon_0^{5}(\delta y,\varphi')}\cos[m(\varphi-\varphi')]
=\frac{(\varphi')^5}{c_8^{5/2}|\varphi'|^5}\cos(m \varphi')
+O(\delta y,\varphi),
%\quad\quad\text{\P S330}
\end{equation}
%~~~~~~~~~~~~~~~~~~~~~~~~~~~~~~~~~~~~~~~~~~~~~~~~~~~~~~~~~~~~~~~~~~~~~~
where we used $\epsilon_0(\delta y=0,\varphi')=c_8^{1/2}|\varphi'|$, $c_8$ being the
coefficient of $\delta\varphi^2$ in $\epsilon_0^2$.
The integral of the leading-order term over domain (ii) vanishes from symmetry,
since this term is anti-symmetric in $\varphi'$ while the integration domain is
symmetric. The integral
over the $O(\delta y,\varphi)$ terms clearly vanishes at the limit
$\delta y,\varphi\to 0$. We conclude, then, that the entire integral in
Eq.\ (\ref{S320}) vanishes at the limit $x\to z$.

We have shown that all contributions to $\tilde{\cal L}^{\alpha m}$ vanish.
Hence $\tilde{\cal L}^{\alpha m}=0$.
Importantly, the vanishing of $\tilde{\cal L}^{\alpha m}$ does not depend on
the direction from which the limit $x\to z$ is taken.

It is instructive to review the above analysis from the point of view
of standard Fourier theory. The quantity $\tilde{\cal L}^{\alpha m}$
is, formally, the Fourier transform of $F^{\alpha}_{\rm dir}-F^{\alpha}_{\rm P}$
(viewed at a function of $\varphi$ for fixed $y$), summed over $\pm m$ for given
$m$, and evaluated at $x\to z$. Eq.\ (\ref{S300}) implies that
$F^{\alpha}_{\rm dir}-F^{\alpha}_{\rm P}$ is a pointwise continuous function
of $\varphi$ for any $y$ (including $y=0$). Standard Fourier theory
[see, e.g., Theorem 4.2 of Ref.\ \cite{James})] tells us that the Fourier series
of such a function, evaluated at some point $\varphi=\varphi_0$, converges to
the {\em average} of the two one-sided values of the function at $\varphi_0$
(even if $\varphi_0$ is a step discontinuity). It is easy to show, based
on Eq.\ (\ref{S300}), that the two-sided average of
$F^{\alpha}_{\rm dir}-F^{\alpha}_{\rm P}$ (with respect to $\varphi$) vanishes
at the limit $\delta x\to 0$. This, reassuringly, is consistent with our
finding $\tilde{\cal L}^{\alpha m}=0$.

Having established Eq.\ (\ref{S250}), we now return to our main line of
development, and to Eq.\ (\ref{S240}). We remind that, by construction, the
limit $x\to z$ in this equation is well defined and direction-independent.
The vanishing of $\tilde{\cal L}^{\alpha m}$ therefore implies that each of the modes
$\tilde f_{\rm res}^{\alpha m}(x)$ is continuous:
$\lim_{x\to z}\tilde f_{\rm res}^{\alpha m}(x)=
\tilde f_{\rm res}^{\alpha m}(z)$. Our mode-sum expression for the tail
part of the SF thus reduces to
%~~~~~~~~~~~~~~~~~~~~~~~~~~~~~~~~~~~~~~~~~~~~~~~~~~~~~~~~~~~~~~~~~~~~~~
\begin{equation}\label{S340}
F_{\rm self}^{\alpha}(z)=
\sum_{m=0}^{\infty} \tilde f_{\rm res}^{\alpha m}(z).
%\quad\quad\text{\P S340}
\end{equation}
%~~~~~~~~~~~~~~~~~~~~~~~~~~~~~~~~~~~~~~~~~~~~~~~~~~~~~~~~~~~~~~~~~~~~~~
It now remains only to relate $\tilde f_{\rm res}^{\alpha m}(z)$ to the
numerical variables $\phi^m_{\rm res}$.
From Eqs.\ (\ref{S90}) and (\ref{S200}) we obtain
$F_{\rm res}^{\alpha}
=q\sum_{m=-\infty}^{\infty} \nabla^{\alpha}(\phi^m_{\rm res}e^{i m \varphi})$,
where we have interchanged the derivative and the sum---this is allowed since
$\Phi_{\rm res}$ is continuous and differentiable (see the discussion below)
and hence its Fourier transform converges uniformly. On the other hand, we have
$F_{\rm res}^{\alpha}=\sum_{m=-\infty}^{\infty}f_{\rm res}^{\alpha m}$.
Since both $\nabla^{\alpha}(\phi^m_{\rm res}e^{i m \varphi})$ and
$f_{\rm res}^{\alpha m}$ depend on $\varphi$ only through a factor $e^{im\varphi}$
[see Eq.\ (\ref{S230})], the orthogonality of the Fourier series implies
%~~~~~~~~~~~~~~~~~~~~~~~~~~~~~~~~~~~~~~~~~~~~~~~~~~~~~~~~~~~~~~~~~~~~~~
\begin{equation}\label{S350}
f_{\rm res}^{\alpha m}=q\nabla^{\alpha}(\phi^m_{\rm res}e^{im\varphi}).
%\quad\quad\text{\P S350}
\end{equation}
%~~~~~~~~~~~~~~~~~~~~~~~~~~~~~~~~~~~~~~~~~~~~~~~~~~~~~~~~~~~~~~~~~~~~~~
In terms of the `tilde' variables this becomes [recalling the definition
of $\tilde\phi^m_{\rm res}$ in Eq.\ (\ref{S130})]
$\tilde f_{\rm res}^{\alpha,m=0}=q\nabla^{\alpha}\tilde\phi^{m=0}_{\rm res}$,
and, for $m>0$,
%~~~~~~~~~~~~~~~~~~~~~~~~~~~~~~~~~~~~~~~~~~~~~~~~~~~~~~~~~~~~~~~~~~~~~~
\begin{eqnarray}\label{S360}
\tilde f_{\rm res}^{\alpha m}&=&
q\nabla^{\alpha}\left(\phi^m_{\rm res}e^{im\varphi}+
       \phi^{-m}_{\rm res}e^{-im\varphi}\right)
%\nonumber\\
%&=&
%q\nabla^{\alpha}\left[e^{im\varphi}\phi^m_{\rm res}+
%\left(e^{im\varphi}\phi^{m}_{\rm res}\right)^*\right]
\nonumber\\
&=&
2q \nabla^{\alpha}{\rm Re}(e^{im\varphi}\phi^m_{\rm res})
\nonumber\\
&=&
q \nabla^{\alpha}\tilde\phi^m_{\rm res}.
%\quad\quad\text{\P S360}
\end{eqnarray}
%~~~~~~~~~~~~~~~~~~~~~~~~~~~~~~~~~~~~~~~~~~~~~~~~~~~~~~~~~~~~~~~~~~~~~~
In the second equality we have made use of the symmetry relation
$\phi^{-m}_{\rm res}=(\phi^{m}_{\rm res})^*$, which is a direct consequence of
$\Phi_{\rm res}$ being a real field.
Substituting $\tilde f_{\rm res}^{\alpha m}(z)=
q \nabla^{\alpha}\tilde\phi^m_{\rm res}(z)$ in Eq.\ (\ref{S340}) finally
establishes the mode-sum formula (\ref{S120}).

%From the definition of the $m$ modes $\phi^m_{\rm res}$ in Eq.\ (\ref{S90})
%we see that these modes are formally obtained from the field $\Phi_{\rm res}$
%through a standard inverse-Fourier transform:
%%~~~~~~~~~~~~~~~~~~~~~~~~~~~~~~~~~~~~~~~~~~~~~~~~~~~~~~~~~~~~~~~~~~~~~~
%\begin{equation}\label{S350}
%\phi^m_{\rm res}=\frac{1}{2\pi}\int_{-\pi}^{\pi}\Phi_{\rm res}
%e^{-im\varphi'}d\varphi'.
%\end{equation}
%%~~~~~~~~~~~~~~~~~~~~~~~~~~~~~~~~~~~~~~~~~~~~~~~~~~~~~~~~~~~~~~~~~~~~~~

\subsubsection{Convergence rate of the $m$-mode sum }

In practice, of course, one can only calculate a finite number of terms in
the mode-sum series (\ref{S120}), and so the question of convergence rate
becomes important. To estimate the convergence rate of the series at large
$m$, recall that the individual $m$-mode contributions
$\tilde f_{\rm res}^{\alpha m}(z)$ arise from taking the limit $x\to z$
of continuous functions $\tilde f_{\rm res}^{\alpha m}(x)$, which themselves
are the Fourier components (multiplied by $e^{im\varphi}$ and summed over $\pm m$
for each $m$) of the function $F_{\rm res}^{\alpha}(x)$. The rate of convergence
of the $m$-mode sum is therefore the rate of convergence of the Fourier series
of $F_{\rm res}^{\alpha}(x)$ at $x=z$.

Here we touch upon a delicate issue: From Eq.\ (\ref{S210}) we have
%~~~~~~~~~~~~~~~~~~~~~~~~~~~~~~~~~~~~~~~~~~~~~~~~~~~~~~~~~~~~~~~~~~~~~~
\begin{equation}\label{S370}
F_{\rm res}^{\alpha}(x)=F_{\rm tail}^{\alpha}(x)+
\left[F_{\rm dir}^{\alpha}(x)-F_{\rm P}^{\alpha}(x)\right],
%\quad\quad\text{\P S370}
\end{equation}
%~~~~~~~~~~~~~~~~~~~~~~~~~~~~~~~~~~~~~~~~~~~~~~~~~~~~~~~~~~~~~~~~~~~~~~
and we recall that $F_{\rm tail}^{\alpha}(x)$ is continuous at $z$, but
$F_{\rm dir}^{\alpha}(x)-F_{\rm P}^{\alpha}(x)$ generally has a
direction-dependence discontinuity there [see Eq.\ (\ref{S300})].
Therefore, $F_{\rm res}^{\alpha}(x)$ is generally {\em discontinuous} at $x=z$,
and, in general, it would have a jump discontinuity in the $\varphi$ direction.
Standard theorem in Fourier analysis (see, e.g., Theorem 4.2 of Ref.\
\cite{James}) predicts, in this case, that the sum of Fourier modes
%$\tilde f_{\rm res}^{\alpha m}(x)$ converges
%to the {\em average} of the two one-sided
%values of $F_{\rm res}^{\alpha}(\varphi)$ at $z$ (for any fixed off-particle
%values of $t,r,\theta$), and that the convergence is
converges very slowly: $\tilde f_{\rm res}^{\alpha m}(z)\sim 1/m$ at large $m$.
Fortunately, however, the situation turns out better in our special case:
As we established above, crucially, each $m$-mode of the difference
$F_{\rm dir}^{\alpha}-F_{\rm P}^{\alpha}$ happens to vanish at $x=z$
(upon summation over $\pm m$ for each $m$), which implies
%~~~~~~~~~~~~~~~~~~~~~~~~~~~~~~~~~~~~~~~~~~~~~~~~~~~~~~~~~~~~~~~~~~~~~~
\begin{equation}\label{S380}
\tilde f_{\rm res}^{\alpha m}(z)=\tilde f_{\rm tail}^{\alpha m}(z).
%\quad\quad\text{\P S380}
\end{equation}
%~~~~~~~~~~~~~~~~~~~~~~~~~~~~~~~~~~~~~~~~~~~~~~~~~~~~~~~~~~~~~~~~~~~~~~
In particular, we find that, at the point $x=z$, the convergence rate of the
Fourier sum of $F_{\rm res}^{\alpha}(x)$ is the same as that of the Fourier
sum of $F_{\rm tail}^{\alpha}(x)$.
The latter function is continuous and piece-wise differentiable at $x=z$
($\nabla^{\beta} F_{\rm tail}^{\alpha}$ will generally have a
direction-dependence discontinuity at that point). For such functions, we
expect $\tilde f_{\rm tail}^{\alpha m}(z)\sim 1/m^2$ at large $m$ [see, e.g., Ref.\
\cite{James}, Sec.\ 4.2.8].
Thus, we also expect $\tilde f_{\rm res}^{\alpha m}(z)\sim 1/m^2$,
and expect that the mode-sum (\ref{S120}) will, in general,
converge like $\sim 1/m$.

\subsubsection{Behavior of $\phi_{\rm res}^m$ near the particle}

When devising a numerical integration scheme for the field equation (\ref{S100}),
it is important to know how regular the $m$-mode fields $\phi_{\rm res}^m$
are near the particle. This, for example, influences the numerical convergence
rate of a given finite-difference scheme. We therefore conclude our analysis by
examining the regularity of $\phi_{\rm res}^m$ near the particle.

From Eqs.\ (\ref{S70}) and (\ref{S170}) we have
$\Phi_{\rm res}=\Phi_{\rm tail}+\left(\Phi_{\rm dir}-\Phi_{\rm P}\right)$.
The tail field, as defined in Eq.\ (\ref{S140}), is a continuous function of $x$
near $z$. From Eq.\ (\ref{S280}) we have $\Phi_{\rm dir}-\Phi_{\rm P}\to\text{const}$
as $x\to z$, and so this difference is continuous as well. Hence,
$\Phi_{\rm res}$ is {\em continuous} at $x=z$. It is then easy to show that the
$m$-modes $\phi_{\rm res}^m$ are also continuous at $x=z$.

We next inspect the differentiability of $\phi_{\rm res}^m$. From Eq.\ (\ref{S360}),
$\nabla^{\alpha}\tilde\phi^m_{\rm res}=\tilde f_{\rm res}^{\alpha m}/q$,
where $\tilde f_{\rm res}^{\alpha m}(x)$, as established above, are continuous
at $x=z$. Hence $\tilde\phi^m_{\rm res}$ are differentiable at $x=z$.
However, it is not the real fields $\tilde\phi^m_{\rm res}$ that interest us
here, but rather the (complex-valued) fields $\phi^m_{\rm res}$ which would
normally serve as variables for the numerical integration. To show that
$\phi^m_{\rm res}$, too, are differentiable at $x=z$, we first argue that
the `non-tilded' functions $f_{\rm res}^{\alpha m}$ are continuous there,
just like $\tilde f_{\rm res}^{\alpha m}$. To see this, consider the limit
${\cal L}^{\alpha m}\equiv
\lim_{x\to z}\left(f_{\rm dir}^{\alpha m}-f_{\rm P}^{\alpha m}\right)$,
which can be evaluated in complete analogy to $\tilde {\cal L}^{\alpha m}$, starting
from Eq.\ (\ref{S310}) with the replacements $\alpha_{m}\to 1$ and
$\cos[m(\varphi-\varphi')]\to \exp[im(\varphi-\varphi')]$. The same
bounding arguments used for $\tilde{\cal L}^{\alpha m}$ can apply for
${\cal L}^{\alpha m}$,
merely replacing $|\cos[m(\varphi-\varphi')]|\leq 1$
with $|\exp[im(\varphi-\varphi')]|\leq 1$. We find, once again, that
only the term in Eq.\ (\ref{S330}) may potentially contribute to the
integral at the limit $x\to z$; however, this time that contribution
fails to vanish by symmetry---in fact, it yields
%~~~~~~~~~~~~~~~~~~~~~~~~~~~~~~~~~~~~~~~~~~~~~~~~~~~~~~~~~~~~~~~~~~~~~~
\begin{equation}\label{S390}
{\cal L}^{\alpha m} \equiv \lim_{x\to z}
\left(f_{\rm dir}^{\alpha m}-f_{\rm P}^{\alpha m}\right)
=\frac{a^{\alpha}_{5,0}}{2\pi}\int_{-\pi}^{\pi}
\frac{(\varphi')^5}{c_{8}^{5/2}|\varphi'|^5}e^{-im\varphi'}
d\varphi'=\left\{
\begin{array}{ll}
-\frac{2ia^{\alpha}_{5,0}}{m\pi c_8^{5/2}}, & \text{for odd $m$},\\
0 ,  & \text{for even $m$}.\\
\end{array}
\right.
%\quad\quad\text{\P S390}
\end{equation}
%~~~~~~~~~~~~~~~~~~~~~~~~~~~~~~~~~~~~~~~~~~~~~~~~~~~~~~~~~~~~~~~~~~~~~~
[Note ${\cal L}^{\alpha,m}=-{\cal L}^{\alpha,-m}$ for all $m$, which, of course,
is consistent with our previous finding, $\tilde{\cal L}^{\alpha m}=0$.]
Since the value of ${\cal L}^{\alpha m}$
does not depend on the direction from which the limit $x\to z$ is taken,
we conclude that $f_{\rm dir}^{\alpha m}-f_{\rm P}^{\alpha m}$ is a continuous
function of $x$ near $z$. Recalling $f_{\rm res}^{\alpha m}=
f_{\rm tail}^{\alpha m}+\left(f_{\rm dir}^{\alpha m}-f_{\rm P}^{\alpha m}\right)$,
and that $f_{\rm tail}^{\alpha m}$ is continuous, we deduce that
$f_{\rm res}^{\alpha m}$ is continuous too. This, by virtue of Eq.\ (\ref{S350}),
implies that the numerical $m$-mode variables $\phi^m_{\rm res}$ are
{\em differentiable} at the particle.

We note that the {\em second} derivatives of  $\phi^m_{\rm res}$ are not
necessarily continuous; indeed, they are not necessarily even bounded at the particle:
We have $q\nabla^{\beta}\nabla^{\alpha}\Phi_{\rm res}=
\nabla^{\beta}F_{\rm res}^{\alpha}=\nabla^{\beta}F_{\rm tail}^{\alpha}+
\nabla^{\beta}\left(F_{\rm dir}^{\alpha}-F_{\rm P}^{\alpha}\right)$,
where $\nabla^{\beta}F_{\rm tail}^{\alpha}$ is bounded at the particle,
but $\nabla^{\beta}\left(F_{\rm dir}^{\alpha}-F_{\rm P}^{\alpha}\right)$
generally diverges there like $\sim 1/\epsilon_0$ [recall Eq.\ (\ref{S300})].
Hence the second derivatives $\nabla^{\beta}\nabla^{\alpha}\Phi_{\rm res}$ generally
diverge at the particle, and their $m$-modes may, potentially, show a logarithmic
divergence there (reminiscent of the logarithmic divergence of
$\phi^m$ \cite{Barack:2007jh}).

In summary, when designing a numerical integration scheme for $\phi^m_{\rm res}$,
one should bare in mind that these fields are {\em continuous and differentiable}
(i.e., have continuous first derivatives), but not necessarily twice-differentiable.

\section{Electromagnetic field} \label{EM}
%%%%%%%%%%%%%%%%%%%%%%%%%%%%%%%%%%%%%%%%%%%%%%%%%%%%%%%%%%%%%%%%%%%%%%%%%%%%%%

In this section we replace the scalar charge of the particle with an
electric charge $e$, and formulate an $m$-mode-sum scheme for the EM SF
acting on the particle. Once again, we assume that the particle
is freely moving in a bound orbit around a Kerr black hole, which, neglecting SF
effects, is a geodesic $z(\tau)$ of the Kerr background. We will be interested
in the value of the EM SF at an arbitrary point along this geodesic,
and, as in the scalar case, we shall ignore the gravitational SF.
The meaning of quantities like $g^{\alpha\beta}$, $u^{\alpha}$, $\tau$, $S$,
$\epsilon_0$, and $\delta x^{\alpha}$ shall remain the same as in the scalar case,
but `$F_{\rm X}$' and `$f^m_{\rm X}$' will now refer to the EM force
(we avoid re-labeling the force as `EM' in the interest of notational simplicity).
The form of the $m$-mode-sum scheme in the EM case, and the analysis leading
to it, resemble very closely those of the scalar case. This will allow
us to skip many of the details of the derivation, pointing frequently to results
from Sec.\ \ref{Scalar}.

\subsection{Preliminaries}

Let us denote the full (retarded) vector potential associated with our charge
by $A_{\alpha}$, and assume that $A_{\alpha}$ can be treated as a linear
field over the fixed Kerr background. In the Lorenz gauge, the vector potential
satisfies the sourced wave-like equation
%~~~~~~~~~~~~~~~~~~~~~~~~~~~~~~~~~~~~~~~~~~~~~~~~~~~~~~~~~~~~~~~~~~~~~~
\begin{equation}\label{V10}
\nabla^{\beta}\nabla_{\beta}A_{\alpha}=
-4\pi e \int_{-\infty}^{\infty}\delta^4[x-z(\tau)](-g)^{-1/2} u_{\alpha}(\tau) d\tau
\equiv-4\pi J_{\alpha},
%\quad\quad\text{\P V10}
\end{equation}
%~~~~~~~~~~~~~~~~~~~~~~~~~~~~~~~~~~~~~~~~~~~~~~~~~~~~~~~~~~~~~~~~~~~~~~
where $J_{\alpha}$ is the charge-current density of the particle.
Including SF effects to $O(e^2)$, the equation of motion of the particle
reads \cite{DeWitt:1960fc,Hobbs:1968,Quinn:1996am}
%~~~~~~~~~~~~~~~~~~~~~~~~~~~~~~~~~~~~~~~~~~~~~~~~~~~~~~~~~~~~~~~~~~~~~~
\begin{equation}\label{V20}
\mu u^{\beta}\nabla_{\beta}u^{\alpha}=F_{\rm self}^{\alpha}=
\lim_{x\to z} F_{\rm tail}^{\alpha}(x),
%\quad\quad\text{\P V20}
\end{equation}
%~~~~~~~~~~~~~~~~~~~~~~~~~~~~~~~~~~~~~~~~~~~~~~~~~~~~~~~~~~~~~~~~~~~~~~
where $F_{\rm self}^{\alpha}$ is now the EM SF. In the EM case, the
``tail force'' field is given by
%~~~~~~~~~~~~~~~~~~~~~~~~~~~~~~~~~~~~~~~~~~~~~~~~~~~~~~~~~~~~~~~~~~~~~~
\begin{equation}\label{V30}
F_{\rm tail}^{\alpha}(x)=e^2 k^{\alpha\beta\gamma}(x)
\lim_{\epsilon\to 0^+} \int_{-\infty}^{\tau_-(x)-\epsilon}
\nabla_{\gamma}G^{\rm ret}_{\beta\beta'}[x,z(\tau')]
u^{\beta'}(\tau')d\tau',
%\quad\quad\text{\P V30}
\end{equation}
%~~~~~~~~~~~~~~~~~~~~~~~~~~~~~~~~~~~~~~~~~~~~~~~~~~~~~~~~~~~~~~~~~~~~~~
where the bi-vector $G^{\rm ret}_{\beta\beta'}$ is the retarded Green's
function associated with Eq.\ (\ref{V10}) [with its un-primed (primed) indices
corresponding to its first (second) argument], $\nabla_{\gamma}$ acts on
the first argument of $G^{\rm ret}_{\beta\beta'}$, and $k$ is a tensor
defined at $x$ through an extension of the velocity vector off the worldline:
%~~~~~~~~~~~~~~~~~~~~~~~~~~~~~~~~~~~~~~~~~~~~~~~~~~~~~~~~~~~~~~~~~~~~~~
\begin{equation}\label{V40}
k^{\alpha\beta\gamma}(x)\equiv
g^{\alpha\gamma}(x)u^{\beta}(x)-g^{\alpha\beta}(x)u^{\gamma}(x).
%\quad\quad\text{\P V40}
\end{equation}
%~~~~~~~~~~~~~~~~~~~~~~~~~~~~~~~~~~~~~~~~~~~~~~~~~~~~~~~~~~~~~~~~~~~~~~
The field $u^{\gamma}(x)$ can be any smooth extension of $u^{\gamma}$;
the SF in Eq.\ (\ref{V20}) depends only of the value of $k$ {\em on} the
worldline, and is not sensitive to the choice of extension.

Eq.\ (\ref{V30}) is analogous to Eq.\ (\ref{S30}) in Sec.\ \ref{Scalar},
and the same conclusions can be drawn in the EM case regarding the regularity of
$F_{\rm tail}^{\alpha}(x)$: This field is continuous and
has at least piecewise continuous (bounded) derivatives.

\subsection{$m$-mode scheme: A prescription}

The procedure in the EM case follows very closely that of the scalar case.
For a puncture function we now take\footnote{In what follows we use the label
`P' alternately as upper- or lower-script, and similarly with the labels `res',
`tail', and `dir'. This should not cause any confusion.}
%~~~~~~~~~~~~~~~~~~~~~~~~~~~~~~~~~~~~~~~~~~~~~~~~~~~~~~~~~~~~~~~~~~~~~~
\begin{equation}\label{V50}
A^{\rm P}_{\alpha}(x)=\frac{e}{\epsilon _{\rm P}(x)}
\left(\bar u_{\alpha}+
\bar\Gamma^{\lambda}_{\alpha\beta}\bar u_{\lambda}\delta x^{\beta}\right),
%\quad\quad\text{\P V50}
\end{equation}
%~~~~~~~~~~~~~~~~~~~~~~~~~~~~~~~~~~~~~~~~~~~~~~~~~~~~~~~~~~~~~~~~~~~~~~
where $\epsilon _{\rm P}$ is the same as in Eq.\ (\ref{S60}), and
$\bar u_{\alpha}$ and $\bar\Gamma^{\lambda}_{\alpha\beta}$ are,
respectively, the four velocity and the Kerr connection coefficients at point $\bar z(t)$.
Recall, for given $x$, $\bar z(t)$ is the point along the particle's worldline
intersected by the surface $t=\text{const}$ containing $x$, and
$\delta x^{\beta}=x^{\beta}-\bar z^{\beta}$.
One will note that, for small $\delta x$, the parenthetical factor in
Eq.\ (\ref{V50}) approximates the value of the four-velocity vector
parallel propagated from $\bar z$ to $x$. We remind, however, that the
definition in Eq.\ (\ref{V50}) applies for any $\delta x$, not necessarily
small.

As in the scalar case, we then define the `residual' field $A^{\rm res}_{\alpha}$,
analogous to $\Phi_{\rm res}$, through
%~~~~~~~~~~~~~~~~~~~~~~~~~~~~~~~~~~~~~~~~~~~~~~~~~~~~~~~~~~~~~~~~~~~~~~
\begin{equation} \label{V60}
A^{\rm res}_{\alpha}=A_{\alpha}-A^{\rm P}_{\alpha},
%\quad\quad\text{\P V60}
\end{equation}
%~~~~~~~~~~~~~~~~~~~~~~~~~~~~~~~~~~~~~~~~~~~~~~~~~~~~~~~~~~~~~~~~~~~~~~
and obtain a wave equation for $A^{\rm res}_{\alpha}$ in the form
%~~~~~~~~~~~~~~~~~~~~~~~~~~~~~~~~~~~~~~~~~~~~~~~~~~~~~~~~~~~~~~~~~~~~~~
\begin{equation}\label{V70}
\nabla^{\beta}\nabla_{\beta}A_{\alpha}^{\rm res}=
-4\pi J_{\alpha}-\nabla^{\beta}\nabla_{\beta}A^{\rm P}_{\alpha}
\equiv Z^{\rm res}_{\alpha}.
%\quad\quad\text{\P V70}
\end{equation}
%~~~~~~~~~~~~~~~~~~~~~~~~~~~~~~~~~~~~~~~~~~~~~~~~~~~~~~~~~~~~~~~~~~~~~~
We next formally decompose $A_{\alpha}^{\rm res}(t,r,\theta,\varphi)$ into
modes $A_{\alpha}^{{\rm res},m}(r,\theta,\varphi)e^{im\varphi}$, just like
in Eq.\ (\ref{S90}), and separate the $\varphi$ dependence in
Eq.\ (\ref{V70}) to obtain
%~~~~~~~~~~~~~~~~~~~~~~~~~~~~~~~~~~~~~~~~~~~~~~~~~~~~~~~~~~~~~~~~~~~~~~
\begin{equation}\label{V80}
\Box^{(3)}_{\rm V}A_{\alpha}^{{\rm res},m}=
\frac{1}{2\pi}\int_{-\pi}^{\pi}Z_{\alpha}^{\rm res}e^{-im\varphi'}d\varphi'
\equiv Z_{\alpha}^{{\rm res},m}.
%\quad\quad\text{\P V80}
\end{equation}
%~~~~~~~~~~~~~~~~~~~~~~~~~~~~~~~~~~~~~~~~~~~~~~~~~~~~~~~~~~~~~~~~~~~~~~
Here $\Box^{(3)}_{\rm V}$ is a certain ($m$-dependent) second-order
differential operator, which couples between the various vectorial components
of $A_{\alpha}^{{\rm res},m}$, but not between different $m$-modes.
Eq.\ (\ref{V80}) constitutes a set of four coupled hyperbolic equations,
with a source term $Z_{\alpha}^{{\rm res},m}$ which, just like in the scalar case,
extends away from the particle. Also in full analogy with the scalar case,
the fields $A_{\alpha}^{{\rm res},m}$ are expected to be continuous and
differentiable (have continuous first derivatives) on the worldline.

Once again, we proceed by assuming that solutions $A_{\alpha}^{{\rm res},m}$
to Eq.\ (\ref{V80}), satisfying physical boundary conditions, have been
obtained. The EM SF is then simply calculated through
%~~~~~~~~~~~~~~~~~~~~~~~~~~~~~~~~~~~~~~~~~~~~~~~~~~~~~~~~~~~~~~~~~~~~~~
\begin{equation}\label{V90}
F_{\rm self}^{\alpha}[z(\tau)]=e \left[k^{\alpha\beta\gamma}\sum_{m=0}^{\infty}
\nabla_{\gamma}\tilde A_{\beta}^{{\rm res},m} \right]_{x=z(\tau)},
%\quad\quad\text{\P V90}
\end{equation}
%~~~~~~~~~~~~~~~~~~~~~~~~~~~~~~~~~~~~~~~~~~~~~~~~~~~~~~~~~~~~~~~~~~~~~~
where $\tilde A_{\beta}^{{\rm res},m}(t,r,\theta,\varphi)$ are real fields
constructed from $A_{\beta}^{{\rm res},m}(t,r,\theta)$ through
%~~~~~~~~~~~~~~~~~~~~~~~~~~~~~~~~~~~~~~~~~~~~~~~~~~~~~~~~~~~~~~~~~~~~~~
\begin{equation}\label{V100}
\tilde A_{\beta}^{{\rm res},m}=
2 {\rm Re}\left(A_{\beta}^{{\rm res},m} e^{im\varphi}\right)\ \text{for $m>0$},
\quad\text{and}\quad \tilde A_{\beta}^{{\rm res},m=0}=A_{\beta}^{{\rm res},m=0}.
%\quad\quad\text{\P V100}
\end{equation}
%~~~~~~~~~~~~~~~~~~~~~~~~~~~~~~~~~~~~~~~~~~~~~~~~~~~~~~~~~~~~~~~~~~~~~~
As in the scalar case, the sum over $m$ in Eq.\ (\ref{V90}) is expected to
converge at least as $\sim 1/m$.

\subsection{Analysis}

The derivation of the mode-sum formula (\ref{V90}) is entirely analogous to the
derivation of Eq.\ (\ref{S120}) in the scalar case. The tail part of the vector
potential is given by
%~~~~~~~~~~~~~~~~~~~~~~~~~~~~~~~~~~~~~~~~~~~~~~~~~~~~~~~~~~~~~~~~~~~~~~
\begin{equation}\label{V110}
A^{\rm tail}_{\beta}(x)=e
\lim_{\epsilon\to 0^+} \int_{-\infty}^{\tau_-(x)-\epsilon}
G^{\rm ret}_{\beta\beta'}[x,z(\tau')]u^{\beta'}(\tau')d\tau',
%\quad\quad\text{\P V110}
\end{equation}
%~~~~~~~~~~~~~~~~~~~~~~~~~~~~~~~~~~~~~~~~~~~~~~~~~~~~~~~~~~~~~~~~~~~~~~
which, recalling Eq.\ (\ref{V30}), gives
%~~~~~~~~~~~~~~~~~~~~~~~~~~~~~~~~~~~~~~~~~~~~~~~~~~~~~~~~~~~~~~~~~~~~~~
\begin{equation}\label{V120}
F_{\rm tail}^{\alpha}(x)=e k^{\alpha\beta\gamma}(x)\left[
\nabla_{\gamma}A_{\beta}^{\rm tail}(x)-
e(\nabla_{\gamma}\tau_-)G^{\rm ret}_{\beta\beta'}[x,z(\tau_-^-)]
u^{\beta'}(\tau_-^-)
\right].
%\quad\quad\text{\P V120}
\end{equation}
%~~~~~~~~~~~~~~~~~~~~~~~~~~~~~~~~~~~~~~~~~~~~~~~~~~~~~~~~~~~~~~~~~~~~~~
Here $G^{\rm ret}_{\beta\beta'}[x,z(\tau_-^-)]\equiv \lim_{\epsilon\to 0^+}
G^{\rm ret}_{\beta\beta'}[x,z(\tau_--\epsilon)]$ contains only the smooth,
tail part of the Green's function, which, by Eq.\ (2.59) of \cite{DeWitt:1960fc},
satisfies $\lim_{x\to z}G^{\rm ret}_{\beta\beta'}[x,z(\tau_-^-)]=-R_{\beta\beta'}/2
+g_{\beta\beta'}R/12$. Since the background Ricci tensor $R_{\beta\beta'}$
and scalar curvature $R$ both vanish in Kerr, and since, as mentioned above,
the factor $\nabla^{\alpha}\tau_-$ is bounded at $x\to z$, we find that the
second term in the square brackets in Eq.\ (\ref{V120}) vanishes at this limit.
We hence obtain
%~~~~~~~~~~~~~~~~~~~~~~~~~~~~~~~~~~~~~~~~~~~~~~~~~~~~~~~~~~~~~~~~~~~~~~
\begin{equation}\label{V130}
F_{\rm tail}^{\alpha}=e k^{\alpha\beta\gamma}
\nabla_{\gamma}A_{\beta}^{\rm tail}
\quad\text{for $x\to z$},
%\quad\quad\text{\P V130}
\end{equation}
%~~~~~~~~~~~~~~~~~~~~~~~~~~~~~~~~~~~~~~~~~~~~~~~~~~~~~~~~~~~~~~~~~~~~~~
which is analogous to Eq.\ (\ref{S160}). Following the same line of derivation as
in Eqs.\ (\ref{S170})--(\ref{S200}), we similarly obtain for the direct field
$F_{\rm dir}^{\alpha}=e k^{\alpha\beta\gamma}
\nabla_{\gamma}A_{\beta}^{\rm dir}$ (for $x\to z$),
and for the full field
$F^{\alpha}=e k^{\alpha\beta\gamma}
\nabla_{\gamma}A_{\beta}$.
The last equality holds for all $x$, provided that the same off-worldline
extension of $k^{\alpha\beta\gamma}$ is chosen for both $F_{\rm tail}^{\alpha}$
and $F_{\rm dir}^{\alpha}$---which we assume here.
We then {\em define}, for all $x$,
%~~~~~~~~~~~~~~~~~~~~~~~~~~~~~~~~~~~~~~~~~~~~~~~~~~~~~~~~~~~~~~~~~~~~~~
\begin{equation}\label{V135}
F_{\rm res}^{\alpha}\equiv e k^{\alpha\beta\gamma}
\nabla_{\gamma}A_{\beta}^{\rm res},
\quad\quad
F_{\rm P}^{\alpha}\equiv e k^{\alpha\beta\gamma}
\nabla_{\gamma}A_{\beta}^{\rm P},
%\quad\quad\text{\P V135}
\end{equation}
%~~~~~~~~~~~~~~~~~~~~~~~~~~~~~~~~~~~~~~~~~~~~~~~~~~~~~~~~~~~~~~~~~~~~~~
with $k^{\alpha\beta\gamma}$ extended the same way as in
$F^{\alpha}$.

With all quantities $F_{\rm X}^{\alpha}$ defined as above for the EM case, we now
reproduce all Eqs.\ (\ref{S210})--(\ref{S240}), in their precise scalar-case form
(with $F^{\alpha}_{\rm X}$, $f^{\alpha m}_{\rm X}$ and $\tilde f^{\alpha m}_{\rm X}$
now, of course, referring to the EM field). It should be noted that the value of
each individual $m$-mode $f^{\alpha m}_{\rm X}$ (or $\tilde f^{\alpha m}_{\rm X}$)
will depend on the specific off-worldline extension chosen for $k^{\alpha\beta\gamma}$.
However, the expression for the final SF in (the EM equivalent of) Eq.\ (\ref{S240})
will {\em not} depend on the $k$-extension.

As in the scalar case, the analysis proceeds by showing the validity of
Eq.\ (\ref{S250}), namely, that the difference $\tilde f_{\rm dir}^{\alpha m}
-\tilde f_{\rm P}^{\alpha m}$ vanishes at the limit $x\to z$.
We start with the expression for the asymptotic form of $A^{\rm dir}_{\beta}(x)$
near $z$, as derived by Mino {\it el al.} \cite{Mino:2001mq}. We write it in the form
\cite{Barack:2002bt}
%~~~~~~~~~~~~~~~~~~~~~~~~~~~~~~~~~~~~~~~~~~~~~~~~~~~~~~~~~~~~~~~~~~~~~~
\begin{equation}\label{V140}
A_{\beta}^{\rm dir}(x)=\frac{e\, \hat u_{\beta}(x)}{\epsilon(x)}
+\frac{e\, w_{\beta}(x)}{\epsilon(x)}+{\rm const}
\quad \text{(for $x$ near $z$)},
%\quad\quad\text{\P V140}
\end{equation}
%~~~~~~~~~~~~~~~~~~~~~~~~~~~~~~~~~~~~~~~~~~~~~~~~~~~~~~~~~~~~~~~~~~~~~~
where $\hat u_{\beta}$ is the four-velocity vector parallel propagated from
$\bar z$ to $x$, and $w_{\beta}$ is a certain smooth function of $x$ (and $\bar z$)
which has the local asymptotic form $w_{\beta}=O(\delta x^2)$, but whose exact
form will not be important for us otherwise. For sufficiently small $\delta x$ we
have $\hat u_{\beta}=\bar u_{\beta}+\bar\Gamma^{\lambda}_{\beta\gamma}
\bar u_{\lambda}\delta x^{\gamma}+O(\delta x^2)$, where the remainder is a smooth
function of $x$. Absorbing this remainder in the function $w_{\beta}$, we rewrite
Eq.\ (\ref{V140}) as
%~~~~~~~~~~~~~~~~~~~~~~~~~~~~~~~~~~~~~~~~~~~~~~~~~~~~~~~~~~~~~~~~~~~~~~
\begin{equation}\label{V145}
A_{\beta}^{\rm dir}=
\frac{e}{\epsilon}\left(
\bar u_{\beta}+
\bar\Gamma^{\lambda}_{\beta\gamma}\bar u_{\lambda}\delta x^{\gamma}\right)
+\frac{e\, w_{\beta}}{\epsilon}+{\rm const}
\quad \text{(for $x$ near $z$)}.
%\quad\quad\text{\P V140}
\end{equation}
%~~~~~~~~~~~~~~~~~~~~~~~~~~~~~~~~~~~~~~~~~~~~~~~~~~~~~~~~~~~~~~~~~~~~~~
Now, from the definition of the puncture function in Eq.\ (\ref{V50}) we
obtain, recalling Eq.\ (\ref{S270}),
%~~~~~~~~~~~~~~~~~~~~~~~~~~~~~~~~~~~~~~~~~~~~~~~~~~~~~~~~~~~~~~~~~~~~~~
\begin{equation}\label{V150}
A_{\beta}^{\rm P}=
\frac{e}{\epsilon}\left(
\bar u_{\beta}+
\bar\Gamma^{\lambda}_{\beta\gamma}\bar u_{\lambda}\delta x^{\gamma}\right)
+\frac{e\, \bar u_{\beta}S_2}{2\epsilon^3}
+O(\delta x^2),
%\quad\quad\text{\P V150}
\end{equation}
%~~~~~~~~~~~~~~~~~~~~~~~~~~~~~~~~~~~~~~~~~~~~~~~~~~~~~~~~~~~~~~~~~~~~~~
and so
%~~~~~~~~~~~~~~~~~~~~~~~~~~~~~~~~~~~~~~~~~~~~~~~~~~~~~~~~~~~~~~~~~~~~~~
\begin{eqnarray}\label{V160}
A_{\beta}^{\rm dir}-A_{\beta}^{\rm P}
&=&\frac{e\left(\epsilon^2 w_{\beta}-\bar u_{\beta}S_2/2\right)}{\epsilon^3}
+{\rm const}+O(\delta x^2)
\nonumber\\
&=&\frac{P_{\beta}^{[4]}(\delta x)}{\epsilon_0^3}
+{\rm const}+O(\delta x^2),
%\quad\quad\text{\P V160}
\end{eqnarray}
%~~~~~~~~~~~~~~~~~~~~~~~~~~~~~~~~~~~~~~~~~~~~~~~~~~~~~~~~~~~~~~~~~~~~~~
where $P_{\beta}^{[4]}$ is of homogeneous order $O(\delta x^4)$, and where we
have absorbed the term arising from the difference between $\epsilon$ and $\epsilon_0$
in the $O(\delta x^2)$ term. Using $F_{\rm dir}^{\alpha}-F_{\rm P}^{\alpha}
=e k^{\alpha\beta\gamma}\nabla_{\gamma}(A_{\beta}^{\rm dir}
-A_{\beta}^{\rm P})$ (for $x\to z$), we then obtain, at the limit $x\to z$,
%~~~~~~~~~~~~~~~~~~~~~~~~~~~~~~~~~~~~~~~~~~~~~~~~~~~~~~~~~~~~~~~~~~~~~~
\begin{equation}\label{V170}
F_{\rm dir}^{\alpha}-F_{\rm P}^{\alpha}=
\frac{\hat P^{\alpha}_{[5]}(\delta x)}{\epsilon_0^5},
%\quad\quad\text{\P V170}
\end{equation}
%~~~~~~~~~~~~~~~~~~~~~~~~~~~~~~~~~~~~~~~~~~~~~~~~~~~~~~~~~~~~~~~~~~~~~~
where
$\hat P^{\alpha}_{[5]}\equiv e k^{\alpha\beta\gamma}(\bar z)\left(
\epsilon_0^2\nabla_{\gamma}P_{\beta}^{[4]}-3P_{\beta}^{[4]}\epsilon_0
\nabla_{\gamma}\epsilon_0\right)$ is of homogeneous order $O(\delta x^5)$.
Eq.\ (\ref{V170}) in entirely analogous to Eq.\ (\ref{S300}), and the analysis
proceeds precisely along the lines of the discussion following the latter
Equation, merely replacing $P_{[5]}^{\alpha}\to \hat P^{\alpha}_{[5]}$. The conclusion,
as in the scalar case, is that $\tilde{\cal L}^{\alpha m}=0$, and,
consequently, we deduce that Eq.\ (\ref{S340}) is valid for the EM SF as well.

We note that the validity of the crucial result $\tilde{\cal L}^{\alpha m}=0$ does not
rely on any specific choice of the $k$-extension, since only the value of $k$
on the worldline enters Eq.\ (\ref{V170}). Therefore, Eq.\ (\ref{S340}) is valid
for any (smooth) extension of $k$. To proceed, we introduce a particular extension
of $k$, denoted $\bar k$, which we define through
%~~~~~~~~~~~~~~~~~~~~~~~~~~~~~~~~~~~~~~~~~~~~~~~~~~~~~~~~~~~~~~~~~~~~~~
\begin{equation}\label{V180}
\bar k^{\alpha\beta\gamma}(x)\equiv k^{\alpha\beta\gamma}(\bar z).
%\quad\quad\text{\P V180}
\end{equation}
%~~~~~~~~~~~~~~~~~~~~~~~~~~~~~~~~~~~~~~~~~~~~~~~~~~~~~~~~~~~~~~~~~~~~~~
Note $\bar k^{\alpha\beta\gamma}$ depends on $t$ [through $\bar z(t)$]
but not on $r,\theta,\varphi$. We denote the functions $F_{\rm res}^{\alpha}$
and $f_{\rm res}^{\alpha m}$ associated with the extension $\bar k$ by
$\bar F_{\rm res}^{\alpha}$ and $\bar f_{\rm res}^{\alpha m}$, respectively.
Since the form of Eq.\ (\ref{S340}) is insensitive to the choice of extension,
we may write
%~~~~~~~~~~~~~~~~~~~~~~~~~~~~~~~~~~~~~~~~~~~~~~~~~~~~~~~~~~~~~~~~~~~~~~
\begin{equation}\label{V190}
F_{\rm self}^{\alpha}(z)=
\sum_{m=0}^{\infty} \tilde{\bar f}_{\rm res}^{\alpha m}(z),
%\quad\quad\text{\P V190}
\end{equation}
%~~~~~~~~~~~~~~~~~~~~~~~~~~~~~~~~~~~~~~~~~~~~~~~~~~~~~~~~~~~~~~~~~~~~~~
where, just as in the scalar case, the `tilde' notation indicates folding
$m<0$ modes over to $m>0$:
$\tilde{\bar f}_{\rm res}^{\alpha m}\equiv \bar f_{\rm res}^{\alpha m}+
\bar f_{\rm res}^{\alpha,-m}$ for $m>0$,
with $\tilde{\bar f}_{\rm res}^{\alpha,m=0}\equiv \bar f_{\rm res}^{\alpha,m=0}$.

It remains to relate the residual force modes $\tilde{\bar f}_{\rm res}^{\alpha m}(z)$
to the numerical variables $A_{\beta}^{{\rm res},m}$.
From Eq.\ (\ref{V135}) we have, choosing the $\bar k$ extension,
$\bar F_{\rm res}^{\alpha}
=e \sum_{m=-\infty}^{\infty} \bar k^{\alpha\beta\gamma}\nabla_{\gamma}
(A_{\beta}^{{\rm res},m}e^{i m \varphi})$. On the other hand,
$\bar F_{\rm res}^{\alpha}=\sum_{m=-\infty}^{\infty} {\bar f}_{\rm res}^{\alpha m}$.
In both sums, each of the $m$-terms depends on $\varphi$ only through a
factor $e^{im\varphi}$, from which we deduce that
%~~~~~~~~~~~~~~~~~~~~~~~~~~~~~~~~~~~~~~~~~~~~~~~~~~~~~~~~~~~~~~~~~~~~~~
\begin{equation}\label{V200}
\bar f_{\rm res}^{\alpha m}=e \bar k^{\alpha\beta\gamma}\nabla_{\gamma}
(A_{\beta}^{{\rm res},m}e^{i m \varphi}).
%\quad\quad\text{\P V200}
\end{equation}
%~~~~~~~~~~~~~~~~~~~~~~~~~~~~~~~~~~~~~~~~~~~~~~~~~~~~~~~~~~~~~~~~~~~~~~
In terms of the `tilde' variables this becomes (recalling
$A_{\beta}^{\rm res}$ is a real field)
%~~~~~~~~~~~~~~~~~~~~~~~~~~~~~~~~~~~~~~~~~~~~~~~~~~~~~~~~~~~~~~~~~~~~~~
\begin{equation}\label{V210}
\tilde{\bar f}_{\rm res}^{\alpha m}=
e \bar k^{\alpha\beta\gamma}\nabla_{\gamma} \tilde A_{\beta}^{{\rm res},m}
%\quad\quad\text{\P V210}
\end{equation}
%~~~~~~~~~~~~~~~~~~~~~~~~~~~~~~~~~~~~~~~~~~~~~~~~~~~~~~~~~~~~~~~~~~~~~~
for all $m\geq 0$. Finally, taking the limit $x\to z$, we may remove the bar
symbol from $\bar k$, since all extensions yield the same value of $k$ on the
worldline. Substituting $\tilde{\bar f}_{\rm res}^{\alpha m}(z)=
e k^{\alpha\beta\gamma}(z)\nabla_{\gamma} \tilde A_{\beta}^{{\rm res},m}(z)$
in Eq.\ (\ref{V190}) establishes the mode-sum formula for the EM case,
Eq.\ (\ref{V90}).

The analysis of the convergence rate of the $m$-mode sum and the regularity of
$A_{\beta}^{{\rm res},m}$ near the particle is entirely analogous to the analysis
in the scalar case, and we shall not reproduce it here. We state the main
results:
(i) The mode sum in Eq.\ (\ref{V90}) is expected to converge at least like
$\sim 1/m$.
(ii) The variables $A_{\beta}^{{\rm res},m}$ are {\em continuous and differentiable}
(i.e., have continuous first derivatives) on the worldline.
(iii) The variables $A_{\beta}^{{\rm res},m}$  are not necessarily
twice-differentiable on the particle's worldline.

%%%%%%%%%%%%%%%%%%%%%%%%%%%%%%%%%%%%%%%%%%%%%%%%%%%%%%%%%%%%%%%%%%%%%%%%%%%%%%
\section{Gravitational perturbations}\label{grav}
%%%%%%%%%%%%%%%%%%%%%%%%%%%%%%%%%%%%%%%%%%%%%%%%%%%%%%%%%%%%%%%%%%%%%%%%%%%%%%

In this section we consider the gravitational SF acting on a test particle
of mass $\mu$, which is moving freely in a bound orbit around a Kerr black hole.
At the limit $\mu\to 0$ the orbit is a geodesic of the Kerr background, again
denoted $z(\tau)$, and we shall prescribe an $m$-mode scheme analogous to the
above for calculating the gravitational SF at an arbitrary point along this
geodesic. Once again we retain the notation for `$F_{\rm X}$' and `$f^m_{\rm
X}$', but these will now refer to the {\em gravitational} force.

The physical nature of the gravitational SF is quite different from that of the
EM (or scalar) forces: It is gauge dependent, and its physical interpretation
is somewhat more subtle. This point is discussed in Ref.\ \cite{Barack:2001ph},
where, in particular, a general gauge-transformation law for the SF is derived.
Our $m$-mode calculation scheme will be formulated within the {\em Lorenz} gauge
(see below)---just like the fundamental formulation in Refs.\ \cite{Mino:1996nk}
and \cite{Quinn:1996am} and like the standard mode-sum scheme. Practically,
this means that the input for our new $m$-mode-sum formula will be the metric
perturbation in the Lorenz gauge, and the output will be the Lorenz-gauge
gravitational SF. The usefulness of the scheme stems from the recent advent of
computational tools for direct integration of the perturbation equations
in the Lorenz gauge \cite{Barack:2005nr,Barack:2007tm}. It should be stressed
that the our mode-sum formula is not at all guaranteed to maintain its form
in gauges other than Lorenz's.

\subsection{Preliminaries}

Let $h_{\alpha\beta}$ denote the full (retarded) linear metric perturbation
associated with our particle. Then define the ``trace-reversed'' perturbation,
$\Psi_{\alpha\beta}\equiv h_{\alpha\beta}-\frac{1}{2}g_{\alpha\beta}h$,
where $g_{\alpha\beta}$ is the background (Kerr) metric, and
$h\equiv h_{\alpha}{}^{\!\alpha}$ is the trace of $h_{\alpha\beta}$.
We assume here that $h_{\alpha\beta}$ is given in the Lorenz gauge, i.e.,
it satisfies the condition $\Psi_{\alpha\beta}{}^{\!;\beta}=0$. Then
the perturbation satisfies
%~~~~~~~~~~~~~~~~~~~~~~~~~~~~~~~~~~~~~~~~~~~~~~~~~~~~~~~~~~~~~~~~~~~~~~~
\begin{equation}\label{G10}
\nabla^{\gamma}\nabla_{\gamma}\Psi_{\alpha\beta}
+2R^{\mu}{}_{\alpha}{}^{\nu}{}_{\beta}\Psi_{\mu\nu}
=-16\pi \mu \int_{-\infty}^{\infty}
\delta^4[x-z(\tau)](-g)^{-1/2}u_{\alpha}u_{\beta}\,d\tau
\equiv-16\pi T_{\alpha\beta},
%\quad\quad\text{\P G10}
\end{equation}
%~~~~~~~~~~~~~~~~~~~~~~~~~~~~~~~~~~~~~~~~~~~~~~~~~~~~~~~~~~~~~~~~~~~~~~~
where $T_{\alpha\beta}$ is the energy-momentum tensor associated with the
particle, and $R^{\mu}{}_{\alpha}{}^{\nu}{}_{\beta}$ is the Riemann tensor
of the Kerr background. The equation of motion of the particle, including $O(\mu^2)$
gravitational SF effects, has precisely the same form as Eq.\ (\ref{V20})
above, with the gravitational ``tail force'' field given by
\cite{Mino:1996nk,Quinn:1996am}
%~~~~~~~~~~~~~~~~~~~~~~~~~~~~~~~~~~~~~~~~~~~~~~~~~~~~~~~~~~~~~~~~~~~~~~
\begin{equation}\label{G20}
F_{\rm tail}^{\alpha}(x)=\mu^2 k^{\alpha\beta\gamma\delta}(x)
\lim_{\epsilon\to 0^+} \int_{-\infty}^{\tau_-(x)-\epsilon}
\nabla_{\delta}G^{\rm ret}_{\beta\gamma\beta'\gamma'}[x,z(\tau')]
u^{\beta'}(\tau')u^{\gamma'}(\tau')d\tau'.
%\quad\quad\text{\P G20}
\end{equation}
%~~~~~~~~~~~~~~~~~~~~~~~~~~~~~~~~~~~~~~~~~~~~~~~~~~~~~~~~~~~~~~~~~~~~~~
Here the bi-tensor $G^{\rm ret}_{\beta\gamma\beta'\gamma'}$ is the retarded Green's
function associated with Eq.\ (\ref{G10}), $\nabla_{\delta}$ acts on
its first argument, and
%~~~~~~~~~~~~~~~~~~~~~~~~~~~~~~~~~~~~~~~~~~~~~~~~~~~~~~~~~~~~~~~~~~~~~~
\begin{equation}\label{G30}
k^{\alpha\beta\gamma\delta}(x)=
         g^{\alpha\delta}u^{\beta}u^{\gamma}/2
        -g^{\alpha\beta}u^{\gamma}u^{\delta}
        -u^{\alpha}u^{\beta}u^{\gamma}u^{\delta}/2
        +u^{\alpha}g^{\beta\gamma}u^{\delta}/4
        +g^{\alpha\delta}g^{\beta\gamma}/4,
%\quad\quad\text{\P G30}
\end{equation}
%~~~~~~~~~~~~~~~~~~~~~~~~~~~~~~~~~~~~~~~~~~~~~~~~~~~~~~~~~~~~~~~~~~~~~~
where both $g^{\alpha\delta}$ and $u^{\beta}$ are evaluated at $x$, the latter
through a smooth extension of the four-velocity off the worldline, just as in
the EM case. In full analogy with the scalar and EM cases, $F_{\rm tail}^{\alpha}(x)$
is continuous on the worldline, and has at least piecewise continuous (bounded)
derivatives there.

\subsection{$m$-mode scheme: A prescription}

%In close analogy with the scalar and EM cases,
Introduce the puncture function
%~~~~~~~~~~~~~~~~~~~~~~~~~~~~~~~~~~~~~~~~~~~~~~~~~~~~~~~~~~~~~~~~~~~~~~
\begin{equation}\label{G40}
\Psi^{\rm P}_{\alpha\beta}(x)=\frac{4\mu}{\epsilon _{\rm P}(x)}
\left[\bar u_{\alpha}\bar u_{\beta}+
\left(\bar\Gamma^{\lambda}_{\alpha\gamma}\bar u_{\beta}+
\bar\Gamma^{\lambda}_{\beta\gamma}\bar u_{\alpha}\right)
\bar u_{\lambda}\delta x^{\gamma}\right],
%\quad\quad\text{\P G40}
\end{equation}
%~~~~~~~~~~~~~~~~~~~~~~~~~~~~~~~~~~~~~~~~~~~~~~~~~~~~~~~~~~~~~~~~~~~~~~
where $\epsilon _{\rm P}(x)$, $\bar u_{\alpha}(t)$ and
$\bar\Gamma^{\lambda}_{\alpha\gamma}(t)$ are the same as in
Eq.\ (\ref{V50}). Then define the residual field through
%~~~~~~~~~~~~~~~~~~~~~~~~~~~~~~~~~~~~~~~~~~~~~~~~~~~~~~~~~~~~~~~~~~~~~~
\begin{equation} \label{G50}
\Psi^{\rm res}_{\alpha\beta}=\Psi_{\alpha\beta}-\Psi^{\rm P}_{\alpha\beta},
%\quad\quad\text{\P G50}
\end{equation}
%~~~~~~~~~~~~~~~~~~~~~~~~~~~~~~~~~~~~~~~~~~~~~~~~~~~~~~~~~~~~~~~~~~~~~~
and obtain a wave equation for $\Psi^{\rm res}_{\alpha\beta}$ in the form
%~~~~~~~~~~~~~~~~~~~~~~~~~~~~~~~~~~~~~~~~~~~~~~~~~~~~~~~~~~~~~~~~~~~~~~
\begin{equation}\label{G60}
\nabla^{\gamma}\nabla_{\gamma}\Psi_{\alpha\beta}^{\rm res}
+2R^{\mu}{}_{\alpha}{}^{\nu}{}_{\beta}\Psi_{\mu\nu}^{\rm res}=
-16\pi T_{\alpha\beta}-\nabla^{\gamma}\nabla_{\gamma}\Psi^{\rm P}_{\alpha\beta}
-2R^{\mu}{}_{\alpha}{}^{\nu}{}_{\beta}\Psi_{\mu\nu}^{\rm P}
\equiv Z^{\rm res}_{\alpha\beta}.
%\quad\quad\text{\P G60}
\end{equation}
%~~~~~~~~~~~~~~~~~~~~~~~~~~~~~~~~~~~~~~~~~~~~~~~~~~~~~~~~~~~~~~~~~~~~~~
Next formally decompose $\Psi_{\alpha\beta}^{\rm res}(t,r,\theta,\varphi)$
into $m$-modes $\psi_{\alpha\beta}^{\rm res}(r,\theta,\varphi)e^{im\varphi}$,
just like in Eq.\ (\ref{S90}), and separate the $\varphi$ dependence in
Eq.\ (\ref{G60}) to obtain
%~~~~~~~~~~~~~~~~~~~~~~~~~~~~~~~~~~~~~~~~~~~~~~~~~~~~~~~~~~~~~~~~~~~~~~
\begin{equation}\label{G70}
\Box^{(3)}_{\rm G}\psi_{\alpha\beta}^{{\rm res},m}=
\frac{1}{2\pi}\int_{-\pi}^{\pi}Z_{\alpha\beta}^{\rm res}e^{-im\varphi'}d\varphi'
\equiv Z_{\alpha\beta}^{{\rm res},m},
%\quad\quad\text{\P G70}
\end{equation}
%~~~~~~~~~~~~~~~~~~~~~~~~~~~~~~~~~~~~~~~~~~~~~~~~~~~~~~~~~~~~~~~~~~~~~~
where $\Box^{(3)}_{\rm G}$ is a certain second-order differential operator (depending
on $m$), which couples between the various tensorial components of
$\psi_{\alpha\beta}^{{\rm res},m}$, but not between different $m$-modes.
Equation (\ref{G70}) constitutes a set of 10 coupled hyperbolic equations for the
10 components of $\psi_{\alpha\beta}^{{\rm res},m}$, with an extended source term.
The modes $\psi_{\alpha\beta}^{{\rm res},m}$ are, by the above construction,
continuous and differentiable on the worldline.

Assuming now that we have at hand solutions $\psi_{\alpha\beta}^{{\rm res},m}$ to
Eq.\ (\ref{G70}), satisfying ``physical'' boundary conditions (in the sense
discussed in Sec.\ \ref{Scalar}), the gravitational
SF is given by the simple formula
%~~~~~~~~~~~~~~~~~~~~~~~~~~~~~~~~~~~~~~~~~~~~~~~~~~~~~~~~~~~~~~~~~~~~~~
\begin{equation}\label{G80}
F_{\rm self}^{\alpha}[z(\tau)]=\mu \left[k^{\alpha\beta\gamma\delta}
\sum_{m=0}^{\infty}
\nabla_{\delta}\tilde\psi_{\beta\gamma}^{{\rm res},m} \right]_{x=z(\tau)},
%\quad\quad\text{\P G80}
\end{equation}
%~~~~~~~~~~~~~~~~~~~~~~~~~~~~~~~~~~~~~~~~~~~~~~~~~~~~~~~~~~~~~~~~~~~~~~
where $\tilde\psi_{\beta\gamma}^{{\rm res},m}(t,r,\theta,\varphi)$ are real fields
constructed from the complex fields $\psi_{\beta\gamma}^{{\rm res},m}(t,r,\theta)$
through
%~~~~~~~~~~~~~~~~~~~~~~~~~~~~~~~~~~~~~~~~~~~~~~~~~~~~~~~~~~~~~~~~~~~~~~
\begin{equation}\label{G90}
\tilde \psi_{\beta\gamma}^{{\rm res},m}=
2 {\rm Re}\left(\psi_{\beta\gamma}^{{\rm res},m} e^{im\varphi}\right)\ \text{for $m>0$},
\quad\text{and}\quad \tilde \psi_{\beta\gamma}^{{\rm res},m=0}=
\psi_{\beta\gamma}^{{\rm res},m=0}.
%\quad\quad\text{\P G90}
\end{equation}
%~~~~~~~~~~~~~~~~~~~~~~~~~~~~~~~~~~~~~~~~~~~~~~~~~~~~~~~~~~~~~~~~~~~~~~
As in the scalar and EM cases, the sum over $m$ in Eq.\ (\ref{G80}) is expected to
converge at least as $\sim 1/m$.

Below we explain the derivation of the mode-sum formula (\ref{G80}),
referring to the scalar/EM cases for many of the details.

\subsection{Analysis}

We start with the formal expression for the tail part of the trace-reversed metric
perturbation,
%~~~~~~~~~~~~~~~~~~~~~~~~~~~~~~~~~~~~~~~~~~~~~~~~~~~~~~~~~~~~~~~~~~~~~~
\begin{equation}\label{G100}
\Psi^{\rm tail}_{\beta\gamma}(x)=\mu
\lim_{\epsilon\to 0^+} \int_{-\infty}^{\tau_-(x)-\epsilon}
G^{\rm ret}_{\beta\beta'\gamma\gamma'}[x,z(\tau')]u^{\beta'}(\tau')
u^{\gamma'}(\tau')d\tau'.
%\quad\quad\text{\P G100}
\end{equation}
%~~~~~~~~~~~~~~~~~~~~~~~~~~~~~~~~~~~~~~~~~~~~~~~~~~~~~~~~~~~~~~~~~~~~~~
Comparing with Eq.\ (\ref{G20}) we find
%~~~~~~~~~~~~~~~~~~~~~~~~~~~~~~~~~~~~~~~~~~~~~~~~~~~~~~~~~~~~~~~~~~~~~~
\begin{equation}\label{G110}
F_{\rm tail}^{\alpha}(x)=\mu k^{\alpha\beta\gamma\delta}(x)\left[
\nabla_{\delta}\Psi_{\beta\gamma}^{\rm tail}(x)-
\mu(\nabla_{\delta}\tau_-)G^{\rm ret}_{\beta\beta'\gamma\gamma'}[x,z(\tau_-^-)]
u^{\beta'}(\tau_-^-)u^{\gamma'}(\tau_-^-)
\right],
%\quad\quad\text{\P G110}
\end{equation}
%~~~~~~~~~~~~~~~~~~~~~~~~~~~~~~~~~~~~~~~~~~~~~~~~~~~~~~~~~~~~~~~~~~~~~~
where $G^{\rm ret}_{\beta\beta'\gamma\gamma'}[x,z(\tau_-^-)]\equiv
\lim_{\epsilon\to 0^+} G^{\rm ret}_{\beta\beta'\gamma\gamma'}[x,z(\tau_--\epsilon)]$ contains
only the smooth, tail part of the Green's function. Unlike in the scalar and EM cases,
here the factor $\propto G^{\rm ret}$ generally does not vanish at the limit $x\to z$:
From Eq.\ (2.11) of \cite{Mino:1996nk} we have
$\lim_{x\to z}G^{\rm ret}_{\beta\beta'\gamma\gamma'}[x,z(\tau_-^-)]=
R_{\beta\gamma\beta'\gamma'}+R_{\beta\gamma'\beta'\gamma}$.
However, it may be readily verified that the contraction $k^{\alpha\beta\gamma\delta}
G^{\rm ret}_{\beta\beta'\gamma\gamma'}u^{\beta'}u^{\gamma'}$ in Eq.\ (\ref{G110}) does
vanish at $x\to z$, for any value of $\alpha$ and $\delta$. (Some of the terms in this
contraction vanish by virtue of the symmetry of the Riemann tensor, and the others due
to the Ricci-flatness of the Kerr spacetime). Thus,
%~~~~~~~~~~~~~~~~~~~~~~~~~~~~~~~~~~~~~~~~~~~~~~~~~~~~~~~~~~~~~~~~~~~~~~
\begin{equation}\label{G120}
F_{\rm tail}^{\alpha}=\mu k^{\alpha\beta\gamma\delta}
\nabla_{\delta}\Psi_{\beta\gamma}^{\rm tail}
\quad\text{for $x\to z$},
%\quad\quad\text{\P G120}
\end{equation}
%~~~~~~~~~~~~~~~~~~~~~~~~~~~~~~~~~~~~~~~~~~~~~~~~~~~~~~~~~~~~~~~~~~~~~~
in analogy with Eqs.\ (\ref{S160}) and (\ref{V130}). We similarly obtain
$F_{\rm dir}^{\alpha}=\mu k^{\alpha\beta\gamma\delta}\nabla_{\delta}
\Psi_{\beta\gamma}^{\rm dir}$ (for $x\to z$), and
$F^{\alpha}=\mu k^{\alpha\beta\gamma\delta} \nabla_{\delta}\Psi_{\beta\gamma}$
(for all $x$), and for all $x$ define
$F_{\rm res}^{\alpha}\equiv \mu k^{\alpha\beta\gamma\delta} \nabla_{\delta}
\Psi^{\rm res}_{\beta\gamma}$ and $F_{\rm P}^{\alpha}\equiv \mu k^{\alpha\beta\gamma\delta}
\nabla_{\delta} \Psi^{\rm P}_{\beta\gamma}$, with the same off-worldline extension of
$k^{\alpha\beta\gamma\delta}$ chosen in all cases. Once again we write the SF
as in Eq.\ (\ref{S240}), and proceed to show that the limit $\tilde{\cal L}^{\alpha m}$,
defined in Eq.\ (\ref{S250}), is null in the gravitational case too.

The form of the direct part of the Lorenz-gauge perturbation was derived by
Mino {\it el al.} \cite{Mino:2001mq}. It can be written as \cite{Barack:2002bt}
%~~~~~~~~~~~~~~~~~~~~~~~~~~~~~~~~~~~~~~~~~~~~~~~~~~~~~~~~~~~~~~~~~~~~~~
\begin{equation}\label{G130}
\Psi_{\beta\gamma}^{\rm dir}(x)=\frac{4\mu\, \hat u_{\beta}(x)\hat u_{\gamma}(x)}{\epsilon(x)}
+\frac{\mu\, w_{\beta\gamma}(x)}{\epsilon(x)}+{\rm const}
\quad \text{(for $x$ near $z$)},
%\quad\quad\text{\P G130}
\end{equation}
%~~~~~~~~~~~~~~~~~~~~~~~~~~~~~~~~~~~~~~~~~~~~~~~~~~~~~~~~~~~~~~~~~~~~~~
where $w_{\beta\gamma}$ is a certain smooth function of $x$ (and $\bar z$) which has
the local asymptotic form $w_{\beta\gamma}=O(\delta x^2)$. [$\hat u_{\beta}$, recall, is the
four-velocity parallel propagated from $\bar z(t)$ to $x$]. As in the EM case, we
replace
$\hat u_{\beta}\to \bar u_{\beta}+\bar\Gamma^{\lambda}_{\beta\delta}
\bar u_{\lambda}\delta x^{\delta}$ in Eq.\ (\ref{G130}), absorbing the (smooth)
$O(\delta x^2)$ error within the function $w_{\beta\gamma}$. We obtain
%~~~~~~~~~~~~~~~~~~~~~~~~~~~~~~~~~~~~~~~~~~~~~~~~~~~~~~~~~~~~~~~~~~~~~~
\begin{equation}\label{G135}
\Psi_{\beta\gamma}^{\rm dir}=\frac{4\mu}{\epsilon}
\left[\bar u_{\beta}\bar u_{\gamma}+
\left(\bar\Gamma^{\lambda}_{\beta\delta}\bar u_{\gamma}+
\bar\Gamma^{\lambda}_{\gamma\delta}\bar u_{\beta}\right)
\bar u_{\lambda}\delta x^{\delta}\right]
+\frac{\mu\, w_{\beta\gamma}}{\epsilon}+{\rm const}
\quad \text{(for $x$ near $z$)}.
%\quad\quad\text{\P G130}
\end{equation}
%~~~~~~~~~~~~~~~~~~~~~~~~~~~~~~~~~~~~~~~~~~~~~~~~~~~~~~~~~~~~~~~~~~~~~~
We wish to compare this local asymptotic form with that of the puncture function
defined in Eq.\ (\ref{G40}). The latter reads, in analogy with Eq.\ (\ref{V150})
of the EM case,
%~~~~~~~~~~~~~~~~~~~~~~~~~~~~~~~~~~~~~~~~~~~~~~~~~~~~~~~~~~~~~~~~~~~~~~
\begin{equation}\label{G137}
\Psi_{\beta\gamma}^{\rm P}=\frac{4\mu}{\epsilon}
\left[\bar u_{\beta}\bar u_{\gamma}+
\left(\bar\Gamma^{\lambda}_{\beta\delta}\bar u_{\gamma}+
\bar\Gamma^{\lambda}_{\gamma\delta}\bar u_{\beta}\right)
\bar u_{\lambda}\delta x^{\delta}\right]
+\frac{2\mu \bar u_{\beta}\bar u_{\gamma}S_2}{\epsilon^3} +O(\delta x^2).
%\quad\quad\text{\P G130}
\end{equation}
%~~~~~~~~~~~~~~~~~~~~~~~~~~~~~~~~~~~~~~~~~~~~~~~~~~~~~~~~~~~~~~~~~~~~~~
Thus,
$\Psi_{\beta\gamma}^{\rm dir}-\Psi_{\beta\gamma}^{\rm P}=
\mu\epsilon^{-3}\left(\epsilon^2 w_{\beta\gamma}-
\bar u_{\beta}\bar u_{\gamma}S_2/2\right)
+{\rm const}+O(\delta x^2)$, which has the form
$\epsilon_0^{-3}P_{\beta\gamma}^{[4]}(\delta x)+{\rm const}+O(\delta x^2)$,
where $P_{\beta\gamma}^{[4]}$ is a certain smooth function of $x$ (and $\bar z$),
of homogeneous order $O(\delta x^4)$. Using $F_{\rm dir}^{\alpha}-F_{\rm P}^{\alpha}
=\mu k^{\alpha\beta\gamma\delta}\nabla_{\delta}(\Psi_{\beta\gamma}^{\rm dir}
-\Psi_{\beta\gamma}^{\rm P})$ (for $x\to z$), we obtain, at the limit $x\to z$,
%~~~~~~~~~~~~~~~~~~~~~~~~~~~~~~~~~~~~~~~~~~~~~~~~~~~~~~~~~~~~~~~~~~~~~~
\begin{equation}\label{G140}
F_{\rm dir}^{\alpha}-F_{\rm P}^{\alpha}=
\frac{\tilde P^{\alpha}_{[5]}(\delta x)}{\epsilon_0^5},
%\quad\quad\text{\P G140}
\end{equation}
%~~~~~~~~~~~~~~~~~~~~~~~~~~~~~~~~~~~~~~~~~~~~~~~~~~~~~~~~~~~~~~~~~~~~~~
where
$\tilde P^{\alpha}_{[5]}\equiv e k^{\alpha\beta\gamma\delta}(\bar z)\left(
\epsilon_0^2\nabla_{\delta}P_{\beta\gamma}^{[4]}-3P_{\beta\gamma}^{[4]}\epsilon_0
\nabla_{\delta}\epsilon_0\right)$ is a smooth function of $x$ (and $\bar z$), of
homogeneous order $O(\delta x^5)$. This, once again, has precisely the form of
the equivalent scalar-case expression, Eq.\ (\ref{S300}), and a similar analysis
(merely replacing $P_{[5]}^{\alpha}\to \tilde P^{\alpha}_{[5]}$) then leads to
the conclusion $\tilde{\cal L}^{\alpha m}=0$. We find that Eq.\ (\ref{S340}) is
valid in the gravitational case too.

The derivation proceeds just as in the EM case: We introduce the ``fixed contravariant
components'' $k$-extension,
$\bar k^{\alpha\beta\gamma\delta}(x)\equiv k^{\alpha\beta\gamma\delta}(\bar z)$,
for which
$\bar f_{\rm res}^{\alpha m}=\mu \bar k^{\alpha\beta\gamma\delta}\nabla_{\delta}
(\psi_{\beta\gamma}^{{\rm res},m}e^{i m \varphi})$. In terms of the `tilde' variables
defined in Eq.\ (\ref{G90}) this becomes (recalling $\Psi_{\beta\gamma}^{\rm res}$
is a real field) $\tilde{\bar f}_{\rm res}^{\alpha m}=\mu \bar k^{\alpha\beta\gamma\delta}
\nabla_{\delta}\tilde \psi_{\beta\gamma}^{{\rm res},m}$, and so
$\tilde{\bar f}_{\rm res}^{\alpha m}(z)=\mu k^{\alpha\beta\gamma\delta}(z)
\nabla_{\delta}\tilde \psi_{\beta\gamma}^{{\rm res},m}(z)$, where we have removed the bar
off $k$ since all extensions coincide at $x=z$. Finally, choosing the $\bar k$ extension
in Eq.\ (\ref{S340}) and substituting for $\tilde{\bar f}_{\rm res}^{\alpha m}(z)$ from
the last expression, we arrive at the mode-sum formula (\ref{G80}).

Once again, the analysis of the convergence rate of the $m$-mode sum and the regularity of
the residual field replicates the scalar-case analysis of Sec.\ \ref{Scalar}, so we merely
re-state the results for the gravitational case:
(i) The mode sum in Eq.\ (\ref{G80}) is expected to converge at least like $\sim 1/m$.
(ii) The variables $\psi_{\beta\gamma}^{{\rm res},m}$ are {\em continuous and differentiable}
on the worldline, but are not necessarily twice-differentiable there.

%%%%%%%%%%%%%%%%%%%%%%%%%%%%%%%%%%%%%%%%%%%%%%%%%%%%%%%%%%%%%%%%%%%%%%%%%%%%%%
\section{Summary and concluding remarks}
%%%%%%%%%%%%%%%%%%%%%%%%%%%%%%%%%%%%%%%%%%%%%%%%%%%%%%%%%%%%%%%%%%%%%%%%%%%%%%

Equations (\ref{S120}), (\ref{V90}) and (\ref{G80}) prescribe the construction of the
scalar, EM, and gravitational SFs, respectively, within our new $m$-mode-sum scheme.
In each case, the raw input for the SF formula are the $m$-modes of the residual function
($\phi^m_{\rm res}$, $A^{{\rm res},m}_{\beta}$, or $\psi^{{\rm res},m}_{\beta\gamma}$), which
are to be obtained through numerical time-evolution of a `punctured' version of the field
equations. Once these residual modes are calculated, they require no further regularization:
The SF is given as a simple sum of (certain combinations of) the derivatives of these
modes, evaluated at the particle's location.

The most computationally-involved stage in the implementation of the scheme is, of course,
the calculation of the $m$-modes of the residual function, which requires numerical
evolution in 2+1D. The feasibility of such calculations was demonstrated in Ref.\
\cite{Barack:2007jh} for the test case of a scalar field in Schwarzschild. Since
the numerical method developed in Ref.\ \cite{Barack:2007jh} does not rely on the
spherical symmetry of the background spacetime, it is directly applicable to a Kerr
background (work to extend the existing scalar code to the Kerr case is underway
\cite{prep}). We envisage applying a similar numerical method to evolve the coupled
set of Lorenz-gauge metric perturbations, but this will require further development.

Working in 2+1D, and calculating the SF using the new $m$-mode scheme, would offer
significant practical advantages, especially in the Kerr case. First, the standard $\ell$-mode
regularization procedure involves the projection of the spheroidal-harmonics components
into a basis of spherical harmonics, which greatly complicates the calculation.
This complication is spared within the new $m$-mode scheme, which avoids
the multipole decomposition altogether. Second, in the standard $\ell$-mode scheme
the numerical variables (i.e, the $\ell,m$ modes of the full fields) require regularization,
which reduces their numerical accuracy: the full modes must be calculated with great initial
accuracy in order for the final, regularized SF to be only moderately accurate (this point
is elaborated on in, e.g., \cite{Barack:2007tm}). In contrast, the numerical variables in
the $m$-mode scheme (i.e, the $m$-modes of the residual fields) require no regularization,
which loosens the numerical accuracy requirements. Third, in the $\ell$-mode scheme one has
to analyze separately the contribution to the mode sum from the ``non-radiative'' low
multipoles ($\ell=0,1$ in the Schwarzschild case); in the Kerr case, merely identifying this
contribution is not a trivial task. The new $m$-mode scheme, on the other hand, requires
no special treatment of the low multipoles---their contribution is automatically
contained within each of the $m$-modes.

This work focuses on establishing the theoretical grounds for the $m$-mode scheme,
and it does not discuss the practicality of the numerical implementation.
Many of these practical issues are discussed in Ref.\ \cite{Barack:2007jh}.
Here we just mention one such implementation issue, which concerns the freedom in
choosing the puncture function. The particular form selected for $\Phi_{\rm P}$
in Eq.\ (\ref{S60}) [and for $A_{\alpha}^{\rm P}$ and $\Psi_{\alpha\beta}^{\rm P}$
in Eqs.\ (\ref{V50}) and (\ref{G40}), respectively] is, of course, not unique.
Focusing, for the moment, on the scalar case, consider the class of puncture functions
%~~~~~~~~~~~~~~~~~~~~~~~~~~~~~~~~~~~~~~~~~~~~~~~~~~~~~~~~~~~~~~~~~~~~~~
\begin{equation} \label{C10}
\hat\Phi_{\rm P}=\frac{q}{\sqrt{S_0+S_1+\Delta S}},
\end{equation}
%~~~~~~~~~~~~~~~~~~~~~~~~~~~~~~~~~~~~~~~~~~~~~~~~~~~~~~~~~~~~~~~~~~~~~~
where $\Delta S$ is any smooth function of the coordinates which vanishes at the
particle at least as $O(\delta x^4)$. Now imagine that we re-formulate the
$m$-mode scheme by replacing $\Phi_{\rm P}\to \hat\Phi_{\rm P}$ (for a specific,
but arbitrary, choice of $\Delta S$). The analysis in Sec.\ \ref{Analysis} can then
be repeated step by step, with only a slight adjustment: In Eqs.\ (\ref{S270})
and (\ref{S280}), we should replace $S_2\to S_2-\Delta S$. Since $S_2-\Delta S$,
just like $S_2$, is a smooth function of order $O(\delta x^4)$, Eq.\ (\ref{S270})
will maintain its form, with only the explicit form of $P_{[4]}(\delta x)$ being
affected. The rest of the analysis does not depend in any way on the explicit
form of $P_{[4]}$. In particular, we find that the final mode-sum formula
(\ref{S120}) is applicable for any choice of a puncture function within the class
$\hat\Phi_{\rm P}$. In just the same way, we can generalize the puncture schemes in
the EM and gravitational cases too, by introducing generalized classes of puncture
functions obtained by taking
$\epsilon_{\rm P}(x)= \sqrt{S_0+S_1+\Delta S}$ in Eqs.\ (\ref{V50}) and (\ref{G40}).
So long as $\Delta S$ is a smooth function of the coordinates which
vanishes at the particle at least as $O(\delta x^4)$, the form of the mode-sum
formulas in Eqs.\ (\ref{V90}) and (\ref{G80}) will not change.

The above freedom in choosing $\Delta S$ can be exploited in the actual implementation
of the puncture scheme, in order to simplify the Fourier integrations involved.
In Ref.\ \cite{Barack:2007jh}, for instance, we have made the replacement
$\delta\varphi^2\to 2(1-\cos\delta\varphi)$ in the expression for $S_0$
(which amounts, in the case considered there, to specifying a certain non-zero function
$\Delta S$), and this allowed us to calculate analytically all the necessary Fourier
integrals. Other choices may simplify the implementation in other cases. One should be
careful, though, to avoid choices of $\Delta S$ which nullify $S_0+S_1+\Delta S$ at
points other than the particle's location.

Finally, we comment that the theoretical foundation for our new $m$-mode
scheme, herein established based on the `tail' + `direct' decomposition, can
alternatively be formulated based on Detweiler and Whiting's `R' + `S'
decomposition \cite{Detweiler:2002mi}. The mathematical details of the analysis
would be quite similar, leading, of course, to the same final $m$-mode formula
for the SF.

\section*{ACKNOWLEDGEMENTS}

This work was supported by PPARC/STFC through grant number PP/D001110/1.
LB thanks the Albert Einstein Institute for hospitality during the completion
of this work.

\end{document}